\documentclass[twocolumn,nofootinbib,superscriptaddress,showkeys]{revtex4-1}
\pdfoutput=1
\usepackage{latexsym,epsfig,amssymb, amsmath,nicefrac}

\usepackage[utf8]{inputenc}

\usepackage{color}

\usepackage[makeroom]{cancel}
  \usepackage{latexsym}
  \usepackage{epsf}
  \usepackage{amssymb}
  \usepackage{graphicx}
  \usepackage{amsmath}
  \usepackage{amsmath,amssymb,amsthm}
  \usepackage{verbatim}
  \usepackage{hyperref}

\def\bi{\begin{itemize}}
\def\ei{\end{itemize}}
\def\be{\begin{equation}}
\def\ee{\end{equation}}
\newcommand{\bea}{\begin{eqnarray}}
\newcommand{\eea}{\end{eqnarray}}
\def\nn{\nonumber}

\newcommand{\eV}{{\, {\rm eV}}}

\newcommand{\GeV}{{\, {\rm GeV}}}
\newcommand{\TeV}{{\, {\rm TeV}}}

\newcommand{\kahler}{K\"{a}hler }  

\def\lsim{\mathrel{\mathop
  {\hbox{\lower0.5ex\hbox{$\sim$}\kern-0.8em\lower-0.7ex\hbox{$<$}}}}}
\def\gsim{\mathrel{\mathop
  {\hbox{\lower0.5ex\hbox{$\sim$}\kern-0.8em\lower-0.7ex\hbox{$>$}}}}}

\begin{document}

\vspace{1cm}
\title{Thermal Dark Energy}

\author{Edward Hardy}
\email{ehardy@liv.ac.uk}
\affiliation{Department of Mathematical Sciences, University of Liverpool, Mathematical Sciences Building
Liverpool, L69 7ZL, UK }
\author{Susha Parameswaran}
\email{susha@liv.ac.uk}
\affiliation{Department of Mathematical Sciences, University of Liverpool, Mathematical Sciences Building
Liverpool, L69 7ZL, UK }

\keywords{Dark energy, finite temperature effects, string theory hidden sectors}

\begin{abstract}
We present a novel source of dark energy, which is motivated by the prevalence of hidden sectors in string theory models and is consistent with all of the proposed swampland conjectures. Thermal effects hold a light hidden sector scalar at a point in field space that is not a minimum of its zero temperature potential. This leads to an effective `cosmological constant', with an equation of state $w=-1$, despite the scalar's zero temperature potential having only a 4D Minkowski or AdS vacuum. For scalar masses $\lesssim \mu$eV, which could be technically natural via sequestering, there are large regions of phenomenologically viable parameter space such that the induced vacuum energy matches the measured dark energy density. Additionally, in many models a standard cosmological history automatically leads to the scalar having the required initial conditions. We study the possible observational signals of such a model, including at fifth force experiments and through $\Delta N_{\rm eff}$ measurements. Similar dynamics that are active at earlier times could resolve the tension between different measurements of $H_0$ and can lead to a detectable stochastic gravitational wave background.

\end{abstract}	

\maketitle

\section{Introduction}

The microscopic nature of dark energy, which constitutes around 70\% of the energy density in the Universe today and drives its current accelerated expansion, is one of the biggest questions in fundamental physics. 
Although observations \cite{Aghanim:2018eyx} are so far consistent with a non-zero cosmological constant, its value would be some 120 orders of magnitude smaller than naive estimates from quantum field theory (see \cite{Martin:2012bt} for a review).  The most commonly considered alternative to a cosmological constant is a slowly rolling scalar field, known as quintessence. However, in models of quintessence the scalar must not only source an extremely small vacuum energy, but it also has to be extremely light with a mass $\lesssim 10^{-32}~\eV$ and yet evade stringent observational constraints on new fifth forces \cite{Tsujikawa:2013fta}.  Moreover, in the absence of additional dynamics, neither a cosmological constant nor quintessence can resolve the growing tension between direct measurements of today's Hubble constant, $H_0$, and the valued inferred from the cosmic microwave background (CMB) using the $\Lambda$CDM model \cite{Riess:2019cxk, Wong:2019kwg}.  A host of current and upcoming astrophysical and cosmological observations will test the $\Lambda$CDM model, quintessence and alternatives, e.g. by probing its equation of state and time dependence \cite{Tanabashi:2018oca}.

String theory provides a well-motivated theoretical framework in which to explore dark energy, particularly in view of the latter's sensitivity to ultraviolet physics and quantum gravity effects.  Various no-go theorems have shown that string theory constructions of vacua with a positive cosmological constant will be, at best, at the limits of theoretical control given current calculational techniques, as reviewed in \cite{Danielsson:2018ztv}. Meanwhile string models of quintessence suffer from similar challenges (see \cite{Cicoli:2018kdo} for a review and \cite{Chiang:2018jdg, Marsh:2018kub, Han:2018yrk, DAmico:2018mnx, Olguin-Tejo:2018pfq, Emelin:2018igk, Hertzberg:2018suv, vandeBruck:2019vzd, Dimopoulos:2019gpz} for some recent works).  

Recently, the difficulty of proving the existence of metastable de Sitter vacua in string theory has prompted a conjecture that any scalar potential arising from a consistent theory of quantum gravity satisfies either
\be
\frac{\sqrt{\nabla^j V \nabla_j V}}{V} \gtrsim \frac{c}{M_{\rm Pl}} \qquad \text{or} \qquad \frac{\text{min}(\nabla^i \nabla_j V)}{V} \lesssim -\frac{c'}{M_{\rm Pl}^2} ~ \label{E:swamp}
\ee
at all points in field space, where $c, c'$ are some universal order one constants \cite{Garg:2018reu, Ooguri:2018wrx}. If true Eq.~\eqref{E:swamp} would rule out metastable de Sitter (dS) vacua, although unstable dS are possible due to the second condition. It would also put slow roll quintessence models under pressure \cite{Agrawal:2018own}.  There are tantalizing connections between this dS Conjecture and other swampland conjectures, as well as with deeper discussions on the consistency of quantum field theory in dS spacetime \cite{Witten:2001kn, Banks:2012hx, Maltz:2016iaw, Dvali:2018jhn, Palti:2019pca}.

In this paper we propose a new, simple, possible source of dark energy, which is consistent with all of the current swampland conjectures. We call this Thermal Dark Energy. Large classes of string compactifications contain an abundance of hidden sectors, which can have rich internal dynamics and interactions yet be only weakly coupled to the visible sector. Consequently it is plausible that there could be a hidden sector with light degrees of freedom that is still in internal thermal equilibrium in the present day Universe. We show that if this is the case, a hidden sector scalar -- which could be a matter or modulus field -- can be stabilised in a temperature-dependent metastable dS minimum, away from the true Minkowski (or AdS) vacuum, by its interactions with the thermal bath. 
Although the equation of state of the induced dark energy is $w=-1$ up to current times, at some point in the future the system will find its way to the global minimum, once the hidden sector temperature has dropped sufficiently.

Considering observational and model building constraints, we show that the measured dark energy density can be explained for scalar masses $\lesssim 10^{-6}$eV, which is still much heavier than the masses required for quintessence. We will also see that Thermal Dark Energy models can lead to observational signals, e.g. at fifth force experiments or in measurements of $\Delta N_{\rm eff}$, that could be accessible to future searches. Additionally, motivated by the richness of typical string compactifications, we consider the possibility that there could be several hidden sectors, at diverse mass scales, each of which sources a component of dark energy. These could lead to detectable gravitational wave signatures and can provide a scenario for Early Dark Energy \cite{Poulin:2018cxd}, which has been argued to resolve the tension between astrophysical and cosmological measurements of the Hubble parameter.

The idea that finite temperature effects may result in a temperature-dependent vacuum energy that leads to cosmic acceleration in the early Universe has previously been proposed under the name of Thermal Inflation  \cite{Lyth:1995hj,Lyth:1995ka}. These papers, and the subsequent literature, focused on dynamics arising from a `flaton' scalar field in the visible sector, at temperatures around the TeV scale. It was shown that of order 10 e-folds of inflation could be obtained, which would dilute otherwise overabundant stable relics and solve the Cosmological Moduli Problem. The key difference of our work is that we consider dynamics occurring in a hidden sector at a much lower scale, so the Universe is currently in the finite temperature generated metastable vacuum, and the accelerated expansion is just beginning, rather than long since finished.

This paper is organised as follows:  In Section~\ref{sec:mod} we show how finite temperature effects around a Minkowski vacuum can provide a source of dark energy, using a simple example mode for illustration.  In Section~\ref{sec:param} we determine the phenomenologically viable parameter space for two realisations of our scenario.  In Section~\ref{sec:signals} we study the potential observable signals of our models, as well as addressing model building issues and discussing how technically natural hidden sectors may be possible. We close with a discussion of our results and open questions in Section~\ref{sec:discuss}.

\section{Dark energy from a thermalised hidden sector} \label{sec:mod}

Suppose there is a hidden sector that includes a scalar field, $\phi$, with a zero temperature potential that stabilises it at a vacuum expectation value (VEV) $\left<\phi\right>=\phi_1$, where the classical vacuum energy vanishes.  As a simple working example, we choose a Higgs-like quartic potential:
\be \label{eq:vphi0}
V_0(\phi) =\lambda \phi^4 -\frac{m_{\phi}^2}{2} \phi^2 + C \,,
\ee
which is minimised at $\phi_1 = m_{\phi}/\left(2 \sqrt{\lambda} \right)$.  The constant $C$ is fixed to $C=m_{\phi}^4 /\left(16 \lambda \right)$ so that the classical vacuum energy vanishes at this point. We will relax this condition below.

We assume that $\phi$ has Higgs-like interactions with other hidden sector states, such that the masses of these increase if $\phi$ develops a VEV. For example the additional states might be fermions, $\psi^i$,  with interactions $y_i \phi \bar{\psi^i} \psi^i$, or other scalars, $\chi^a$, with interactions $\lambda_a \phi^2 \chi^a \chi^a$. 

If it is to source Thermal Dark Energy, the hidden sector must be in internal thermal equilibrium in the present day Universe. However, the hidden sector temperature $T_{\rm h}$ need not be the same as the temperature of the visible sector (and due to observational constraints typically cannot be). Thermal equilibrium is maintained provided $\Gamma_{\rm I} > H$, where $\Gamma_{\rm I}$ is the typical rate of hidden sector interactions and $H$ is the Hubble parameter. The hidden sector can be in thermal equilibrium even if the $\phi$ quartic coupling $\lambda$ in Eq.~\eqref{eq:vphi0} is small, provided the  interactions between $\phi$ and the additional hidden sector states are sufficiently strong. As we discuss later, the condition for thermalisation is easily satisfied provided that the masses of $\phi$ and the other hidden sector states are smaller than the hidden sector temperature. 

Finite temperature effects -- where the plasma interacts with the homogenous scalar field background which itself determines the masses and interactions of particles in the plasma -- affect the dynamics of $\phi$  (see e.g. \cite{Bellac:2011kqa}). This is encapsulated in the thermal effective potential for $\phi$, which at one loop is given by
\bea \label{eq:thermpot}
V(\phi_c,T_{\rm h}) &=& V_0 \left( \phi_c \right)   \label{E:VT} \\ && + \frac{T_{\rm h}^4}{2\pi^2}\left(-\sum_{\psi^i} n_{\psi^i} \,J_F\left(\frac{m_{\psi^i}^2(\phi_c)}{T_{\rm h}^{2}}\right) \right. \nn \\  && \left. \quad \quad \quad \quad + \sum_{\chi_a}  n_{\chi^a} \,J_B\left(\frac{M_{\chi_a}^2(\phi_c)}{T_{\rm h}^{2}}\right)\right)~, \nn
\eea
where $m_{\psi^i}^2\left(\phi_c\right)$ and $M_{\chi_a}^2\left(\phi_c \right)$ are the masses of the hidden sector fermions and scalars respectively in the homogeneous background of $\phi$ denoted $\left<\phi\right> = \phi_c$, and the sums run over all species in thermal equilibrium. Eq.~\eqref{eq:thermpot} is expressed in terms of the thermal functions
\bea
J_B(x^2) &\equiv & \int_0^\infty dq \, q^2\, \log\left(1-e^{-\sqrt{q^2+x^2}}\right)  \\
J_F(x^2) &\equiv & \int_0^\infty dq \, q^2\, \log\left(1+e^{-\sqrt{q^2+x}^2}\right)~,
\eea
where $n_{\psi^i}$ and $n_{\chi^a}$ are the number of degrees of freedom in the fermion $i$ and the scalar $a$ respectively.\footnote{The one loop thermal potential leads to numerical inaccuracies in some theories, and it can be improved in various ways (see e.g. \cite{Curtin:2016urg}). We have verified that including higher corrections via daisy resummation does not significantly change the allowed parameter space in the models we consider.} Although we use their full expressions when plotting results, it is instructive to consider the high and low temperature behaviours of the thermal functions.\footnote{These can be derived from $J_B(x^2)=  -\sum_{n=1}^{\infty} n^{-2} x^2 K_2 (n\,x)$ and $J_F(x^2) =  -\sum_{n=1}^{\infty} (-1)^n n^{-2} x^2 K_2 (n\,x)$, where $K_2$ is the modified Bessel function of the second kind.}  In the high temperature limit, $|x| \ll 1$,
\bea \label{eq:htexp}
J_B(x^2) &=& -\frac{\pi^4}{45} + \frac{\pi^2}{12}x^2 + \mathcal{O}(x^3) \nn\\
J_F(x^2) &=& \frac{7\pi^4}{360} - \frac{\pi^2}{24}x^2 + \mathcal{O}(x^3)~,
\eea
whereas in the low temperature limit, $|x| \gg 1$,
\bea \label{eq:ltexp}
J_B(x^2) &=& x^2 e^{-x} \sqrt{\frac{\pi}{2x}}\nn\\
J_F(x^2) &=& -x^2 e^{-x} \sqrt{\frac{\pi}{2x}}~. \label{E:lowT}
\eea
For simplicity we assume the hidden states that acquire mass from a $\phi$ VEV are otherwise massless. Then at temperatures much higher than the masses in the thermal bath, $T_{\rm h} \gg m_{\psi^i}(\phi_c), m_{\chi^a}(\phi_c)$, that is $T_{\rm h} \gg y_i \phi_c, \lambda_a \phi_c$, the thermally corrected scalar potential goes as:
\be \label{eq:vlt}
V(\phi,T_{\rm h}) = \lambda \phi^4 -\frac{m_{\phi}^2}{2} \phi^2 + \frac{m_{\phi}^4}{16 \lambda} -a T_{\rm h}^4 + b T_{\rm h}^2 \phi^2 ~.
\ee
The constant $a$ in Eq.~\eqref{eq:vlt} is fixed by the number of light degrees of freedom, while $b$  can be inferred from the preceding expressions and depends on the couplings $y_i$ and $\lambda_a$. The effect of a $\phi$ background on its own mass also contributes to $b$, however we will see later that the quartic self interaction of $\phi$ is suppressed in phenomenologically interesting models so this is negligible.  In the simple case that $\phi$ couples only to a single Dirac fermion with Yukawa coupling $y=1$ we have $b = 1/12$. Meanwhile any states with a mass $m_i\left(\phi_c\right) \gg T_{\rm h}$ do not contribute significantly to $\phi$'s thermal potential, as can be seen from the expansion Eq.~\eqref{eq:ltexp}.

From Eq.~\eqref{eq:vlt}, if 
\be \label{eq:tcond}
T_{\rm h} > \frac{m_{\phi}}{\sqrt{2 b}}~,
\ee
finite temperature effects lead to a minimum at $\phi =0$, since the high temperature approximation to the thermal potential, Eq.~\eqref{eq:vlt}, is automatically valid around this point. The minimum can be either stable or metastable. If Eq.~\eqref{eq:vlt} is also accurate around $\phi$'s zero temperature VEV at $\phi_1=m_{\phi}/\left(2 \sqrt{\lambda} \right)$, $\phi=0$ will be the global minimum of the thermal potential. However, if $\lambda \ll 1$ (so $\phi_1 \gg m_\phi$) the high temperature approximation might break down at large $\phi$ if the induced masses of the hidden sector states become larger than the hidden sector temperature. In this case the thermal contribution to $\phi$'s potential is exponentially suppressed, and the zero temperature potential dominates in this part of field space. As a result there is a minimum close to $\phi=\phi_1$, which is deeper than that at $\phi=0$ if $T_{\rm h}^4 \ll m_{\phi}^4/\lambda$.

\begin{figure}[t!]
\begin{center}
\includegraphics[scale=0.38]{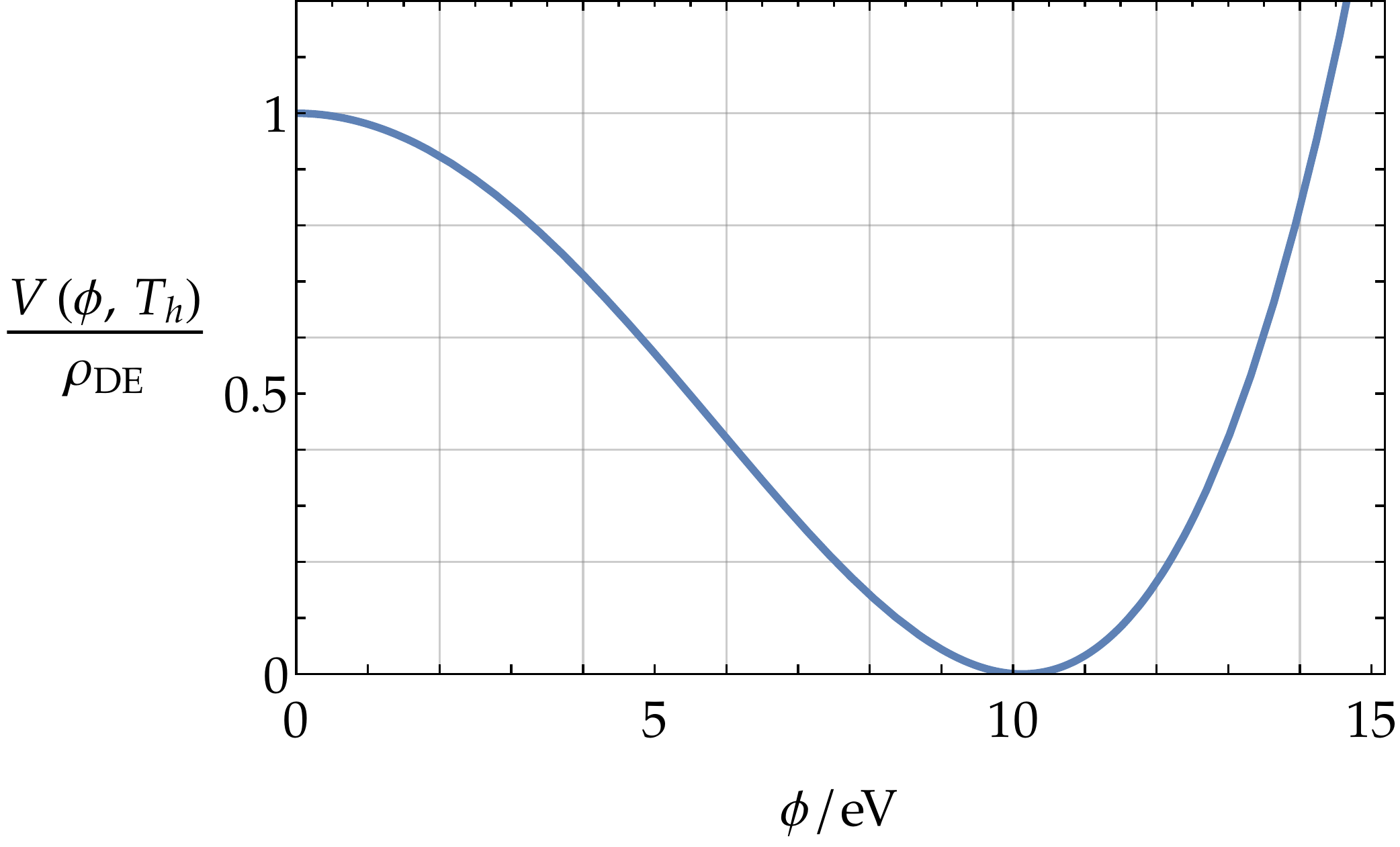}
\end{center}
\caption{The finite temperature scalar potential, defined by Eqs.~\eqref{eq:vphi0} and \eqref{E:VT}, for a model with $m_{\phi}=10^{-6}~\eV$ and $\lambda = 2.44 \times 10^{-15}$, in which $\phi$ is coupled to a single Dirac fermion with a Yukawa coupling $y=1$. The hidden sector temperature is fixed to  $T_{\rm h}= 10^{-4.5}~\eV$. The vertical axis is normalised relative to the present day dark energy density $\rho_{\rm DE} = \left( 0.0023~\eV \right)^4$, and the potential is symmetric around $\phi=0$. \label{F:VT}}
\end{figure}

\begin{figure}[t!]
\begin{center}
\includegraphics[scale=0.415]{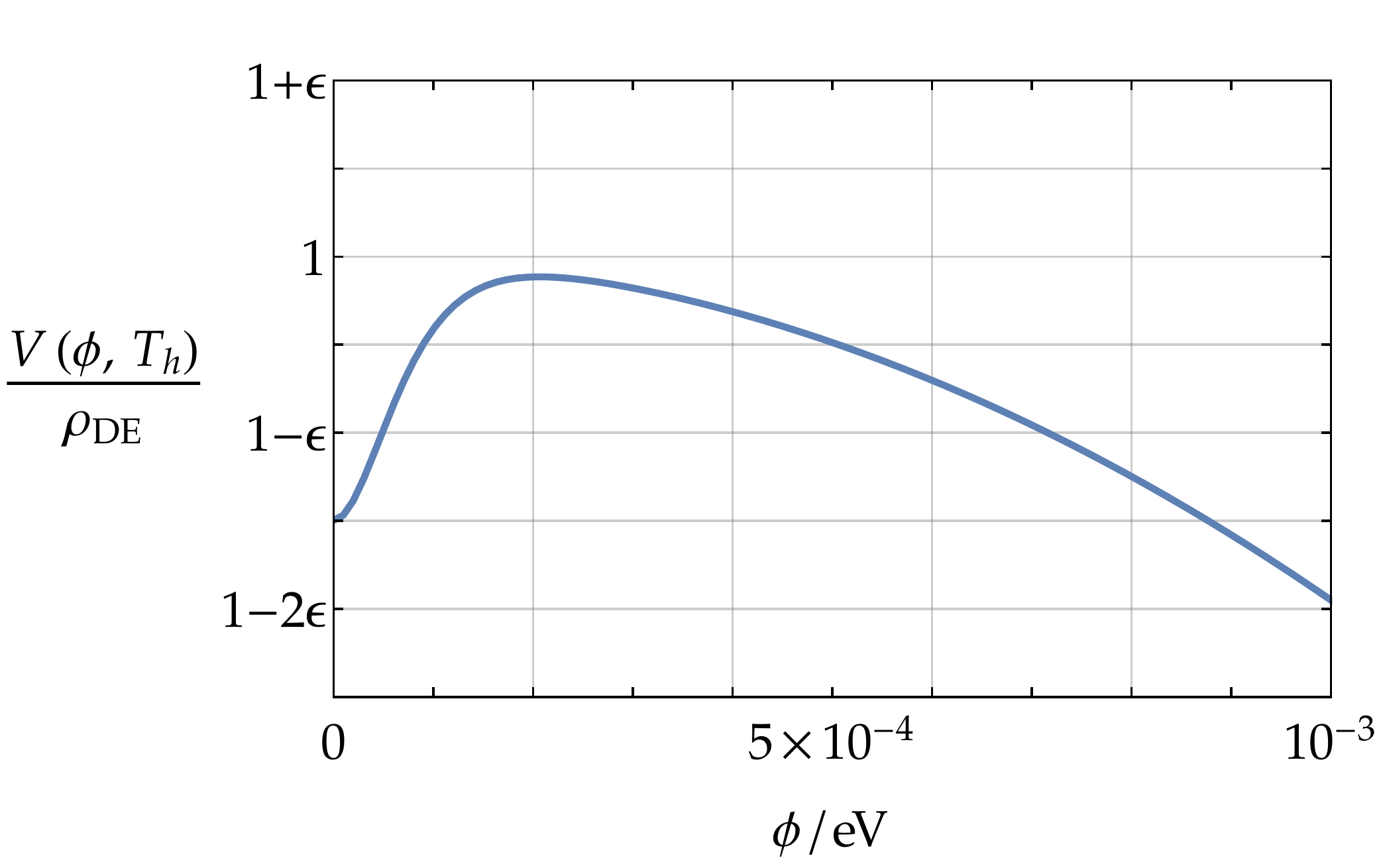}
\end{center}
\caption{Close up of the potential in Figure \ref{F:VT} around $\phi = 0$, showing the metastable minimum due to the thermal potential. The vertical axis is plotted in steps of $\epsilon = 10^{-8}$. \label{F:VTzoom}}
\end{figure}

In Figure~\ref{F:VT} we plot the thermal potential, Eq.~\eqref{eq:thermpot}, of $\phi$ for a model with $m_{\phi} = 10^{-6}~\eV$, $\lambda = 2.4\times 10^{-15}$ and $b=1/12$. The hidden sector temperature $T_{\rm h} = 10^{-4.5}~\eV$ is such that there is a local minimum at the origin, however this is only metastable and the minimum close to $\phi_1$ is much deeper. A plot of the same potential zoomed in around the origin is shown in Figure~\ref{F:VTzoom}. Moving away from $\phi=0$ the potential increases until $\phi \gtrsim T_{\rm h}$ at which point the masses of the hidden sector fermions become larger than the temperature, and the contribution from these states to the thermal potential is suppressed.

If $\phi$ is trapped in the minimum at $\phi=0$, the hidden sector will contribute to the Einstein's equations determining the evolution of the Universe in two ways. First, there is an energy density contained in hidden sector radiation of
\be
\rho_{r {\rm h}} = \frac{\pi^2 g_{\rm h} {T_{\rm h}^0}^4}{30}~,
\ee
where $g_{\rm h}$ is the effective number of hidden sector relativistic degrees of freedom, and ${T_{\rm h}^0}$ is the current hidden sector temperature (and we assume that the hidden sector does not contain any significant abundance of non-relativistic matter).\footnote{The effect of the hidden sector temperature being smaller than then visible sector temperature is not absorbed in our definition of $g_{\rm h}$.} Second, even though the finite temperature effects break Lorentz invariance and do not contribute directly to the observed vacuum energy which sources the Einstein's equations, they can contribute indirectly by shifting the vacuum expectation value of $\phi$ (see \cite{Chung:2011it} for a clear discussion on this point). As a result, $\phi$ will source a contribution to the dark energy density of
\be \label{eq:rhode}
\rho_{\rm DE} = \frac{m_{\phi}^4}{16 \lambda}~.
\ee
It is this contribution to the dark energy density that we will use to explain the accelerated expansion of the Universe.

\section{Viable parameter space}\label{sec:param}

If it is to account for the observed present day dark energy, a hidden sector scalar with potential given in Eq.~\eqref{eq:vphi0} must satisfy various model building and observational constraints. In studying these, we assume that the minimum of the zero temperature potential of the Universe has zero cosmological constant, although there is also viable parameter space in which this is anti-de Sitter (AdS). For definiteness we also suppose that $\phi$ is coupled to a single Dirac fermion with Yukawa coupling $y=1$, however our results are not very sensitive to this choice.

To account for the observed dark energy density we need $\rho_{\rm DE} \simeq (2.3~ {\rm meV})^4$ \cite{Aghanim:2018eyx}. Since the dark energy density is larger than the energy density in radiation in present day Universe,  Eq.~\eqref{eq:rhode} implies 
\be
\frac{m_{\phi}^4}{16 \lambda} > \frac{\pi^2 g_{\rm v}}{30} {T_{\rm v}^0}^4~, 
\ee
where $T_{\rm v}^0 \sim 0.24 ~{\rm meV}$ is the temperature of the visible sector CMB photons today and $g_{\rm v}$ is the effective number of visible sector relativistic degrees of freedom. As well as needing to be subdominant to the dark energy density, the energy density in hidden sector radiation in the present day Universe is strongly constrained by its impact on the effective number of relativistic degrees of freedom at earlier times (as we discuss shortly). Observational bounds require that the present day hidden sector temperature, $T_{\rm h}^0$, is smaller than $T_{\rm v}^0$.

In addition to these conditions on the energy densities, the hidden sector must satisfy Eq.~\eqref{eq:tcond} so that a metastable minimum exists. In our current model all the different requirements can be satisfied simultaneously  if $\lambda \ll 1$, so $m_{\phi} \ll \phi_1$, and also the hidden sector temperature is such that $m_{\phi} \ll T_{\rm h}^0 \ll \phi_1$. Put another way, a small quartic coupling allows the temperature induced mass of $\phi$ at $\phi=0$ to be
\be
b {T_{\rm h}^0}^2 > -\frac{\partial^2 V_0}{\partial \phi^2} ~,
\ee
which creates a local minimum, despite there being a large energy difference between the metastable and global minima
\be
{T_{\rm h}^0}^4 \ll V\left(0\right) - V\left(\phi_1\right) ~.
\ee
The local minimum at $\phi=0$ is automatically metastable relative to the minimum at $\phi_1$ in this regime. 

The requirement of a small coupling, in order for Thermal Dark Energy to dominate over radiation energy in our present model, is mirrored for other forms of zero temperature potential, which generically need a small dimensionless parameter, or a hierarchy in dimensionful scales to be in this regime.\footnote{The absence of such a hierarchy is why the Standard Model Higgs does not lead to a period of Thermal Dark Energy domination when the visible sector temperature is around the electroweak scale.}

It is interesting to note that the condition $C = m_{\phi}^4 /\left(16 \lambda \right)$ in Eq.~\eqref{eq:vphi0}, for the potential's zero temperature minimum to be Minkowski, can be relaxed.  Ensuring that $\phi=0$ is a metastable minimum still requires Eq.~\eqref{eq:tcond}, and the dark energy density is still given by $\rho_{\rm DE}=C$.  Assuming that $C < m_{\phi}^4 /\left(16 \lambda \right)$, so that the zero minimum temperature is AdS, then implies that $\lambda < m_{\phi}^4/\left(16 \rho_{\rm DE} \right) < 4 b^2 T_{\rm h}^4/\left(16 \rho_{\rm DE} \right)$.

\begin{figure}[t!]
\begin{center}
\includegraphics[scale=0.39]{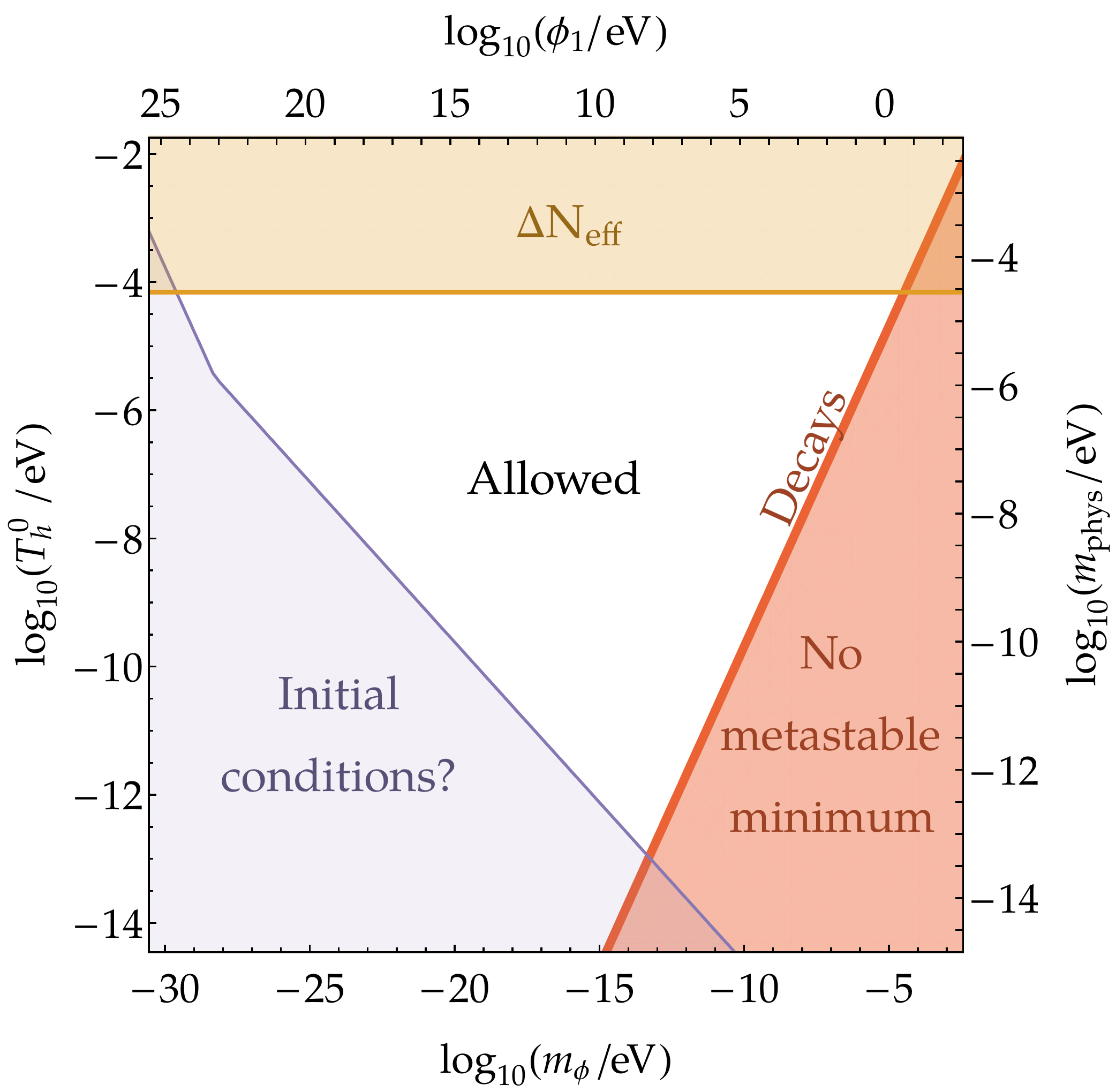}
\end{center}
\caption{The parameter space in which a scalar with zero temperature potential given by Eq.~\eqref{eq:vphi0} coupled to a single Dirac fermion and trapped in a thermally generated metastable minimum can account for the present day dark energy. The results are shown as a function of the Lagrangian parameter $m_{\phi}$, the present day hidden sector temperature $T_{\rm h}^0$, the zero temperature VEV, $\phi_1$, and the physical mass of $\phi$ excitations around the metastable vacuum, $m_{phys}$.  Constraints arise from demanding that a suitable metastable minimum exists, and an additional barely visible region is excluded by requiring that this is sufficiently long lived. The hidden sector temperature $T_{\rm h}^0$ cannot be too large otherwise observational constraints on the effective number of neutrinos are excluded. Outside the region labeled ``Initial conditions?'' the scalar is automatically in the metastable minimum provided the reheating temperature after inflation is sufficiently high. Models inside this region are not excluded but do require an explaination of why $\phi$ is in the metastable minimum.\label{F:cons1}}
\end{figure}

The parameter space in which there is a metastable minimum that can account for the observed dark energy density is plotted in Figure~\ref{F:cons1}, for $C = m_{\phi}^4 /\left(16 \lambda \right)$, as a function of the Lagrangian parameter $m_{\phi}$ and the present day hidden sector temperature $T_{\rm h}^0$. Note that $m_{\phi}$ does not coincide with the physical mass of the scalar around the metastable minimum in the present day Universe, which is instead dominantly sourced by thermal effects. Apart from very close to the stability boundary, the current physical mass is well approximated by
\be \label{eq:mphys}
m_{\rm phys}^2 = 2 b {T_{\rm h}^0}^2 ~.
\ee
Over all of the parameter space of interest, $\phi$'s zero temperature potential satisfies the dS swampland condition Eq.~\eqref{E:swamp}, and the distance in field space between the high and low temperature minima, $\Delta \phi= 2\rho_{\rm DE}^{1/2}/m_{\phi}$, is much smaller than $M_{\rm Pl}$.

Figure~\ref{F:cons1} also shows the other phenomenological constraints, which we now describe. 

\subsubsection{Temperature of the Hidden Sector and $\Delta N_{\rm eff}$}

The energy density in hidden sector particles is constrained by their effect on the measured expansion history of the Universe. For relativistic dark particles this is usually parameterised by the effective number of neutrino species, $N_{\rm eff}$, or the effective number of degrees of freedom, $g_*^{\rm eff}$, defined via (see e.g. \cite{Feng:2008mu, Cheung:2010gj, Berlin:2016gtr})
\bea
g_*^{\rm eff} & \equiv & g_*^{{\rm SM}-\nu} + g_*^\nu \xi_\nu^4 + g_*^{\rm h} \xi_{\rm h}^4 \\
& = & g_*^{{\rm SM}-\nu}  + \frac78 \times 2 \times N_{\rm eff} \times (\xi_\nu^0)^4 \,. \label{E:Neffdef}
\eea
Here $g_*$ counts the number of degrees of freedom, weighted by 1 for bosons and $\frac78$ for fermions, ${\rm SM}-\nu$ refers to the Standard Model (SM) without neutrinos, $\xi_\nu \equiv T_\nu/T_{\rm v}$ (where $T_{\rm v}$ is the temperature of the visible sector photon bath), and 
\be
\xi_{\rm h} \equiv T_{\rm h}/T_{\rm v}
\ee
accounts for the possibility that the hidden sector is cold relative to the visible sector. $\xi_\nu^0$ is $\xi_\nu$ in the Standard Model when neutrino reheating from electron-positron annihilations is neglected, that is $\xi_\nu^0 = (4/11)^{1/3}$ for $T_{\rm v} \lesssim m_e$ and $\xi_\nu^0=1$ for $T_{\rm v} \gtrsim m_e$.  

We assume that the hidden and visible sectors are reheated to different temperatures and that, being only very weakly coupled, interactions between the two sectors do not significantly alter their temperature ratio, $\xi_{\rm h}$.  However, $\xi_{\rm h}$ will still vary throughout the history of the Universe due to separate conservation of the co-moving entropies, $s=g_{s}(T) T^3$, within each sector.  Assuming that only the visible sector number of degrees of freedom change, this leads to the relation 
\be
\xi_{\rm h}(T_2) = \xi_{\rm h}(T_1) \left(\frac{g_{s\,{\rm v}}(T_2)}{g_{s\, {\rm v}}(T_1)}\right)^{1/3} ~.
\ee 
Solving \eqref{E:Neffdef} for $N_{\rm eff}$ we have, with three neutrino flavours,
\be
N_{\rm eff} = 3 \left(\frac{\xi_\nu}{\xi_\nu^0}\right)^4 + \frac47 g_*^{\rm h} \left(\frac{\xi_{\rm h}}{\xi_\nu^0}\right)^4 ~.
\ee  
Observational limits on $N_{\rm eff}$, both from the abundances of light elements and the properties of the CMB, then constrain the hidden sector temperature and effective number of degrees of freedom. Around big bang nucleosynthesis (BBN), $T_\nu = T_\gamma$ and $\xi_\nu^0=\xi_\nu$, and we have
\be
N_{\rm eff} \approx 3 + \frac47 g_*^{\rm h} \xi_{\rm h}^4 ~.
\ee
 Assuming e.g. $g_*^{\rm h}=1+\frac78 4$, as is the case in our illustrative model containing a real scalar and a single Dirac fermion, and using the upper bound on $N_{\rm eff}$, $N_{\rm eff} \lesssim 3.18$ from BBN constraints \cite{Pitrou:2018cgg}, this implies $T_{\rm h} \lesssim 0.51 T_{\rm v}$ at BBN.  Taking into account the change in degrees of freedom from BBN to today in the visible sector, $g_{s\,{\rm v}}=10.75$ to $g_{s\,{\rm v}}=2$, we obtain an upper limit on the current temperature of the hidden sector
\be
T_{\rm h}^0 \lesssim 0.29 T_{\rm v}^0 ~.
\ee
Around the time of recombination and the formation of the CMB, instead:
\be
N_{\rm eff} \approx 3.046 + \frac47 \left(\frac{11}{4}\right)^{4/3} g_*^h \xi_{\rm h}^4 ~.
\ee
Using the upper boundary for CMB constraints on $N_{\rm eff}$, $N_{\rm eff} \lesssim 3.55$ \cite{Aghanim:2018eyx}, this leads to a weaker limit:
\be
T_{\rm h}^0 \lesssim  0.47 T_{\rm v}^0~.
\ee

\subsubsection{Lifetime of the Universe} \label{sec:decays}

As discussed above, if dark energy dominates over the energy in hidden sector radiation the vacuum at $\phi =0$ is automatically only metastable, and it can therefore decay through nucleation of bubbles of the true vacuum. For a model to explain the current dark energy this must occur at a sufficiently slow rate 
\be \label{eq:rate}
\Gamma_{\rm nucl} \ll H_0^4 ~, 
\ee
where $H_0$ is the present day value of the Hubble parameter. Nucleation of bubbles can occur via quantum fluctuations tunneling through the barrier between the minima or thermal fluctuations travelling up the hill. The rate of such events depends on the critical bubble actions $S_4$ or $S_3/T_{\rm h}$ respectively, which lead to nucleation rates that are approximately
\bea \label{eq:rates}
\Gamma_3 &=& T_{\rm h}^4 \left(\frac{S_3}{2\pi T_{\rm h}} \right)^{3/2} e^{-S_3/T_{\rm h}} \\
\Gamma_4 &=& v^4 \left(\frac{S_4}{2\pi}\right)^2 e^{-S_4} ~,
\eea
where $v$ is the width of the barrier that is tunnelled through \cite{Affleck:1980ac,Turner:1992tz}.\footnote{A more precise determination of the prefactor in $\Gamma_4$ is possible \cite{Strumia:1998nf}, however this is unimportant for our purposes.} 

Eq.~\eqref{eq:rate} imposes lower bounds on $S_3/T_{\rm h}^0$ and $S_4$. The value of $S_3/T_{\rm h}^0$ above which the Universe is sufficiently stable explicitly depends on the ratio of the hidden and visible sector temperatures via the prefactor in Eq.~\eqref{eq:rates}, however this is only a logarithmic effect. For example, if $T_{\rm h}^0 = 0.2~ T_{\rm v}^0$, vacuum stability requires that $S_3 /T_{\rm h}^0 \gtrsim 270$, while if $T_{\rm h}^0 = 0.02~ T_{\rm v}^0$ we need $S_3 /T_{\rm h}^0 \gtrsim 260$.

To determine the parameter space in which the metastable minimum is sufficiently long lived, we calculate the critical bubble actions $S_3/T_{\rm h}^0$ and $S_4$ as a function of $m_{\phi}$ and $T_{\rm h}^0$ (having fixed that the model sources the measured dark energy density). This is done via a standard numerical overshoot/ undershoot method, using a modified version of the publicly available code \textsc{CosmoTransitions}\cite{Wainwright:2011kj}. 
Since the distance in field space between the metastable minimum and the global minimum $\phi_1$ is much larger than the barrier height, the field configuration of a critical bubble interpolates between $\phi=0$ and a value on the other side of the barrier $\phi_*$ that satisfies $\phi_*\ll \phi_1$.\footnote{As a result the thin wall approximation \cite{Coleman:1977py} to the bubble actions is extremely inaccurate.}

\begin{figure}[t!]
\begin{center}
\includegraphics[scale=0.4]{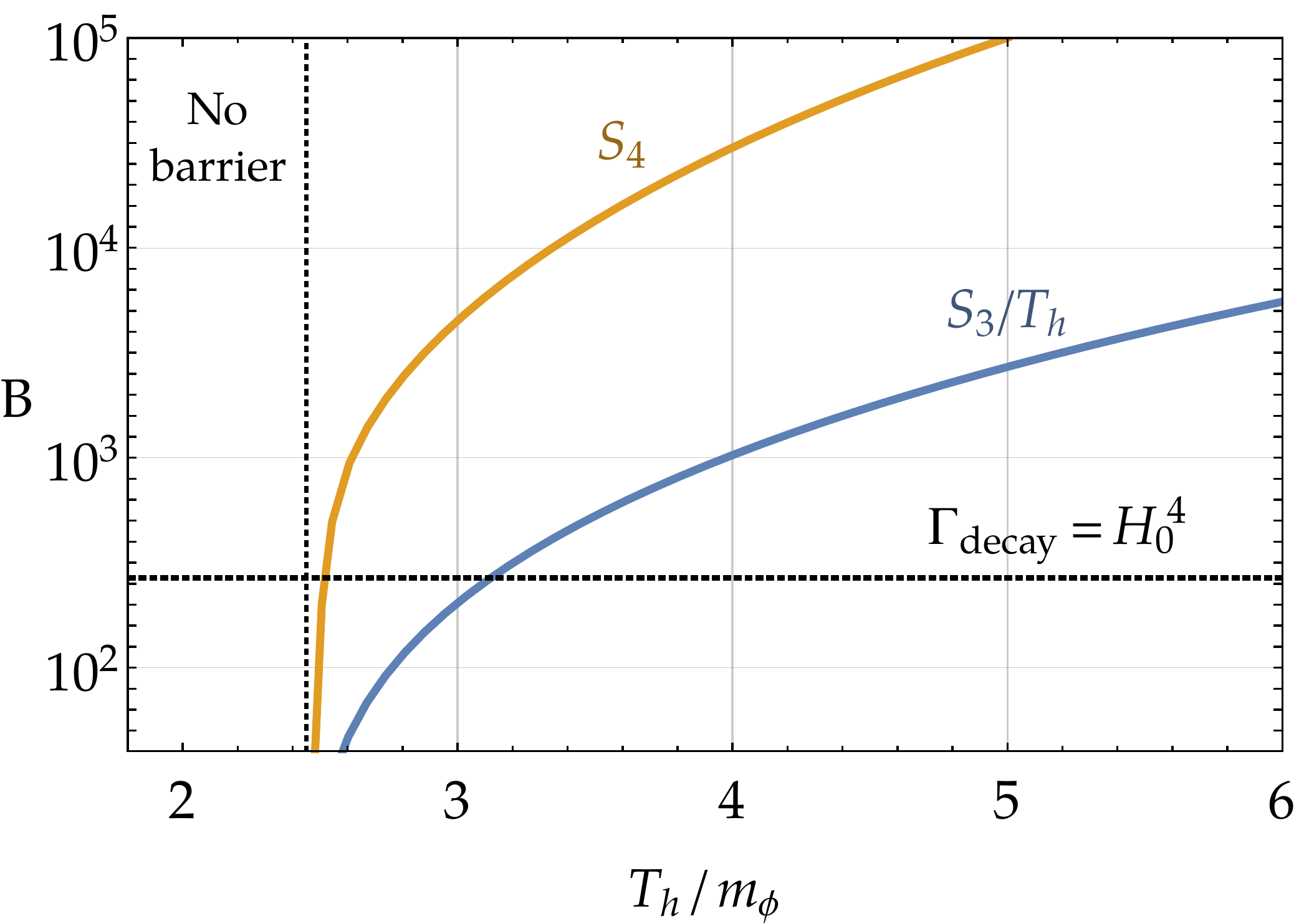}
\end{center}
\caption{The actions $B$ for the formation of critical bubbles by thermal fluctuations $S_3/T_{\rm h}$ and quantum tunnelling $S_4$ as a function of the hidden sector temperature, for a model with $m_{\phi}= 10^{-6}~\eV$ and a quartic coupling such that the metastable minimum sources the measured present day dark energy density. The temperature at which the barrier between the minima disappears, $T=\sqrt{6} m_{\phi}$, and the approximate value of the actions required for vacuum decay to occur within a Hubble time in the present day Universe are also plotted.  \label{F:decay}}
\end{figure}

The critical bubble actions are shown for a model with $m_{\phi} = 10^{-6}~\eV$ in Figure~\ref{F:decay} as a function of the hidden sector temperature. Both actions remain sufficiently large so that vacuum decay is negligible until close to $T_{\rm h} = \sqrt{6} m_{\phi}$, when the metastable minimum disappears.

Similarly to in Figure~\ref{F:decay}, over all of the parameter space of our model vacuum decay remains negligible until shortly before the barrier between minima disappears, and $\Gamma_3$ is exponentially larger than $\Gamma_4$. The critical bubble actions increase quickly as a function of the hidden sector temperature, so if the decay rate is sufficiently small at present times then it was negligible at earlier times as well. Consequently only a thin strip of otherwise allowed models are excluded by the vacuum stability constraint in Figure~\ref{F:cons1}. 

Having discussed vacuum decay, let us consider the rest of the cosmological history. 
Dark energy domination began at a time $t_{\rm i}$ such that $ \rho_{\rm DE} > 3 H_0^2 \Omega_{\rm M}/a\left(t_{\rm i} \right)^3$, where $\Omega_{\rm M}$ is the present day matter density (including dark matter) and $a\left(t\right)$ is the scale factor of the Universe normalised such that $a=1$ today. 

Looking to the future, dark energy domination will end at a time $t_{\rm e}$ such that $T_{\rm h} \approx m_{\phi}$. Since the critical bubble actions $S_3/T_{\rm h}$ and $S_4$ go to zero as the barrier between the minima disappears, the rate of bubble nucleation will grow exponentially immediately prior to this. As a result the Universe will go through a first order phase transition to the zero temperature minimum at this time. Because the barrier disappears at some non-zero hidden sector temperature, the era of accelerated expansion necessarily ends. During the phase transition the dark energy sourced by $\phi$ will be converted to hidden sector radiation or matter (apart from a small fraction that goes into  gravitational waves).

The number of e-folds between the beginning and the end of dark energy domination will be
\bea \label{eq:nefold}
N_{\rm DE} &=& \log\left(\frac{a\left(t_{\rm e}\right)}{a\left(t_{\rm i}\right)}\right) \nonumber\\
&=& \log\left(\frac{T_{\rm h}^0}{m_{\phi}}\right) + 0.65 ~,
\eea
where we used that the number of e-folds between the onset of dark energy domination and today is $N=0.65$.  For example, for $T^{\rm h}_0 = 10^{-4.5}~{\rm eV}$ and $m_{\phi} =10^{-6}~\eV$, we have $N_{\rm DE} = 4.1$. 

\subsubsection{Thermal equilibrium and initial conditions}

We now briefly return to the issue of thermal equilibrium in the hidden sector.  To be maintained at a time $t$, this requires 
\be \label{eq:thermeq}
\Gamma_{\rm I}\left(t\right) \gg H\left(t\right)~, 
\ee
where $\Gamma_{\rm I} = n \left< \sigma v \right>$ is the thermally averaged rate of an interaction that maintains equilibrium, $n$ is the hidden sector number density, $\sigma$ is the cross section, and $v$ is the velocity of the particles (see e.g. \cite{Kolb:1990vq}).  As long as the temperature of the hidden sector is much greater than the masses of the hidden sector particles they are relativistic, $v \sim c$, $m\sim 0$, $n \sim T_{\rm h}^3$ and $\sigma \sim \alpha/T_{\rm h}^2$ for some dimensionless constant $\alpha$, so $\Gamma \sim \alpha T_{\rm h}$. Then Eq.~\eqref{eq:thermeq} was satisfied during the radiation dominated era if 
\be \label{eq:thmcond}
\xi_{\rm h} \gg \frac{1}{\alpha} \frac{T_{\rm v}}{ M_{\rm Pl}} ~.
\ee
Provided $\alpha$ is not tiny, this is easily the case for, say, $T_{\rm v} \lesssim 10^9~\GeV$ over all of the parameter space that we are interested in (and $T_{\rm v}$ may be significantly higher for some regions of parameter space). Similarly it can be seen that equilibrium is subsequently maintained during matter domination, and the present era, as long as $\alpha$ is not minuscule.

\begin{figure}[t!]
\begin{center}
\includegraphics[scale=0.41]{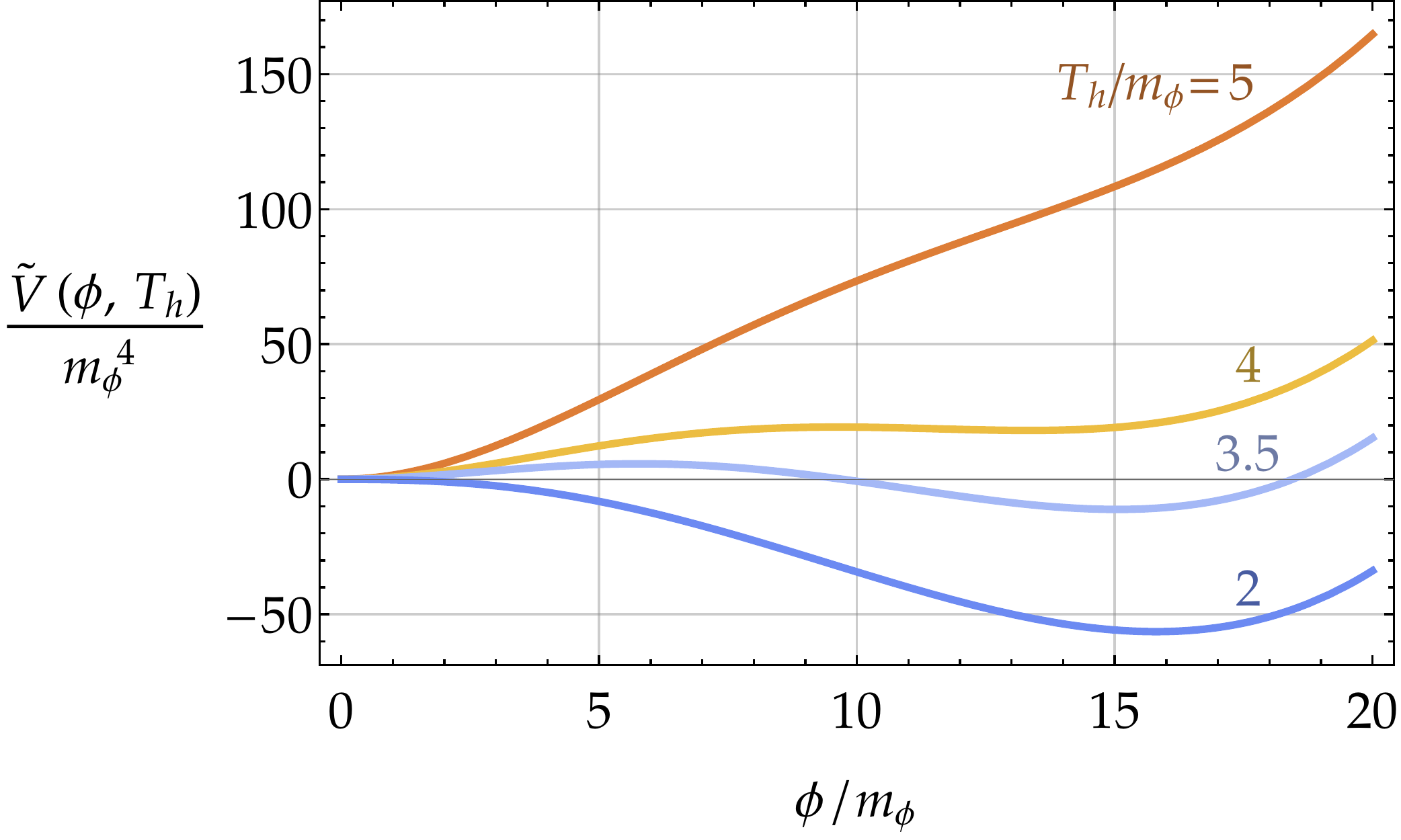}
\end{center}
\caption{The finite temperature scalar potential, defined by Eqs.~\eqref{eq:vphi0} and \eqref{E:VT} and shifted by a temperature dependent constant defined in Eq.~\eqref{eq:shift} for clarity, of a theory with a quartic coupling $\lambda = 10^{-3}$, for different hidden sector temperatures (assuming $\phi$ is coupled to a single Dirac fermion with a Yukawa coupling $y=1$). At high temperatures $\phi=0$ is the only minimum of the potential. As the hidden sector temperature decreases a deeper minimum forms, but $\phi$ remains trapped in the, now metastable, minimum at $\phi=0$.
 \label{F:initcond}}
\end{figure}

It is also natural to ask how the Universe found itself in a minimum that is so energetically disfavoured. This is actually an automatic outcome of a normal cosmological history over the majority of the parameter space of our model. In Figure~\ref{F:initcond} we show $\phi$'s finite temperature potential for different hidden sector temperatures in our illustrative model.\footnote{In this figure we assume a quartic coupling $\lambda =10^{-3}$. Although such a value is too large for the induced Thermal Dark Energy to explain the present day dark energy density in viable models, the key qualitative features of the potential are unaffected and are more conspicuous when plotted.} For clarity, we plot the finite temperature potential shifted by a temperature dependent constant
\be \label{eq:shift}
\tilde{V}\left(\phi,T_{\rm h}\right) = V\left(\phi,T_{\rm h}\right)-V\left(0,T_{\rm h}\right) ~.
\ee
If the temperature of the hidden sector was greater than $\phi_1$ at some time during the history of the Universe, and the hidden sector was in internal thermal equilibrium at this point, $\phi =0$ will have been the global minimum of the potential while the region around $\phi_1$ was unstable. At later times, once the hidden sector temperature has dropped, there will be a deeper minimum close to $\phi_1$. However, as long as the transition rate to the true vacuum is sufficiently slow, $\phi$ remains trapped at $\phi=0$, and it sources Thermal Dark Energy.\footnote{These dynamics are reminiscent of those proposed to occur in some models of supersymmetry breaking \cite{Abel:2006cr}.}

Such a history is not essential for a phenomenologically successful model. However, since it is simple and minimal, we consider the conditions for it to occur. First, the hidden sector must reach a sufficiently high temperature after inflation. Neglecting small effects from the changes in degrees of freedom in the two sectors, this requires that the visible sector reheating temperature satisfies
\be
T_{\rm v}^{\rm RH} \gtrsim \frac{\phi_1}{\xi_{\rm h}} ~,
\ee
which, in a model that sources the present day dark energy density, corresponds to
\be \label{eq:rhcond}
T_{\rm v}^{\rm RH} \gtrsim \frac{\rho_{\rm DE}^{1/2}}{\xi_{\rm h} m_{\phi}} ~.
\ee

Negative searches for primordial gravitational waves bound the scale of inflation $H_{\rm{I}} \lesssim 10^{14}~\GeV$   \cite{Ade:2015tva}, which assuming fast reheating constrains $T_{\rm v}^{\rm RH} \lesssim 10^{16}~\GeV$. Combined with Eq.~\eqref{eq:rhcond} this limits the parameter space of our model.\footnote{Due to the finite time taken for thermalisation after inflation this condition might be too weak \cite{Davidson:2000er}, however this only slightly affects our allowed parameter space.}

The hidden sector must also be in thermal equilibrium at $T_{\rm h} = \phi_1$, so that $\phi$ rolls away from this point in field space. As an estimate of whether this is the case, we suppose that the relevant hidden sector interactions have coupling constants of $\mathcal{O}\left(1\right)$ so they occur at a rate $\Gamma_{\rm I} \sim T_{\rm h}$. Similarly to the derivation of Eq.~\eqref{eq:thmcond}, thermal equilibrium requires
\be \label{eq:eqcond}
m_{\phi} \xi_{\rm h}^2 > \frac{\rho_{\rm DE}^{1/2}}{M_{\rm Pl}} ~.
\ee

The constraints in Eqs.~\eqref{eq:rhcond} and \eqref{eq:eqcond} are plotted in Figure~\ref{F:cons1}. Apart from a small region with $m_{\phi} \lesssim 10^{-28} \eV$ and $T_{\rm h}^0 \sim 10^{-5}~\eV$, Eq.~\eqref{eq:eqcond} is the stronger of the two conditions. Over most of the parameter space $\phi_1$ and therefore the required reheating temperature is not especially large. We also stress that these are not sharp bounds, and indeed these conditions are not needed at all if the cosmological history of the Universe was different.

\subsubsection{An alternative potential} \label{sec:mod2}

A thermal potential can also create a metastable minimum, and shift the vacuum energy density, in theories with zero temperature potentials that differ from Eq.~\eqref{eq:vphi0}. As we will discuss in Section~\ref{sec:signals}, the phenomenological and model building features of a model can vary depending on the form of its zero temperature potential.

As an example, we show that Thermal Dark Energy is possible in a theory with a modulus-like zero temperature potential
\be \label{eq:mod2}
V_0 \left(\phi\right)= \frac{m_{\phi}^2}{2}\left( \phi -\phi_1 \right)^2 ~.
\ee
Despite having a single minimum at $\phi=\phi_1$ and no stationary point at $\phi=0$, after coupling $\phi$ to other hidden sectors states the finite temperature potential can still create a metastable minimum close to $\phi=0$. If such a minimum exists it sources a dark energy density $ \rho_{\rm DE} \simeq \frac12 m_{\phi}^2 \phi_1^2$. This has a chance of dominating the energy density of the Universe (i.e. $\rho_{\rm DE} \gg {T_{\rm v}^0}^4$) if the hidden sector has a hierarchy of scales $m_{\phi} \ll T_{\rm h}^0 \lesssim T_{\rm v}^0 \ll \phi_1$. 

The condition for the thermal potential to generate a vacuum at $\phi \ll \phi_1$ is slightly different to the previous model since the zero temperature potential has a non-zero gradient at the origin. If it is in a region of field space for which the quadratic approximation to the thermal potential is valid, the thermally generated minimum is at
\be \label{eq:min2}
\phi_0 \left(T_{\rm h} \right) = \frac{m_{\phi}^2 \phi_1}{m_{\phi}^2 + 2 b T_{\rm h}^2} ~.
\ee
As before we assume that $b$ is generated by a Yukawa coupling between $\phi$ and an otherwise massless hidden sector Dirac fermion, so $b= y^2/12$. The quadratic approximation is accurate and the minimum in Eq.~\eqref{eq:min2} is self consistent provided $\phi_0\left(T_{\rm h}\right) < y T_{\rm h}$. In the phenomenologically interesting scenario that the induced vacuum energy exceeds the energy in the hidden sector radiation, this leads to a condition on the hidden sector temperature
\be \label{eq:th0mod3}
T_{\rm h}^0 \gtrsim \frac{1}{y}  m_{\phi}^{2/3} \phi_1^{1/3} ~.
\ee
Requiring that the present day measured dark energy density is induced, we have
\be
\phi_1 = \sqrt{2}\rho_{\rm DE}^{1/2}/m_{\phi} ~,
\ee
and Eq.~\eqref{eq:th0mod3} becomes
\be \label{eq:mod2cond}
T_{\rm h}^0 \gtrsim \frac{1}{y} m_{\phi}^{1/3} \rho_{\rm DE}^{1/6} ~.
\ee
It can be shown that no metastable minimum exists if Eq.~\eqref{eq:mod2cond} is violated by a factor of more than $\mathcal{O}\left(1\right)$. 

\begin{figure}[t!]
\begin{center}
\includegraphics[scale=0.39]{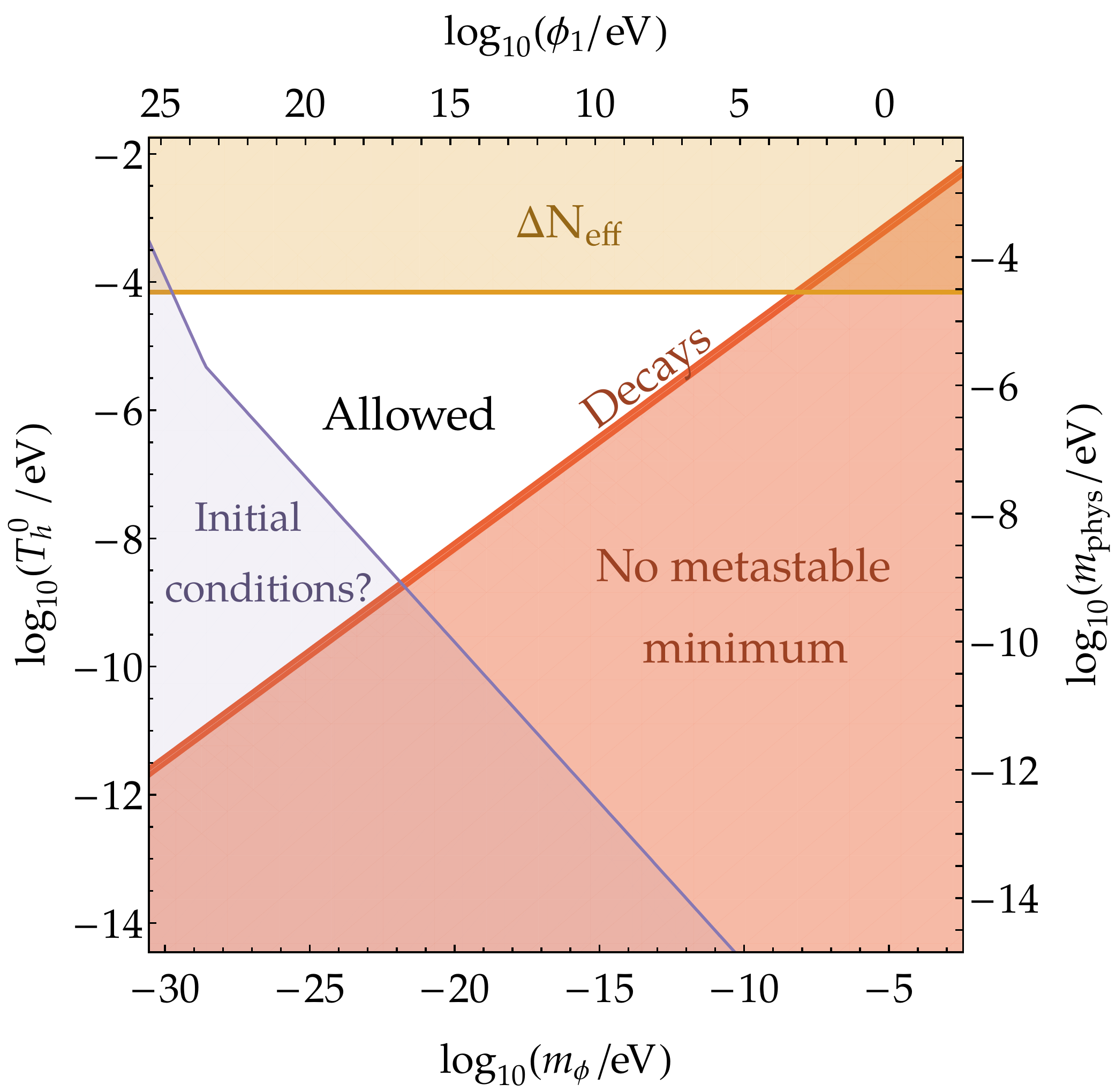}
\end{center}
\caption{Constraints on a model with a zero temperature potential given by Eq.~\eqref{eq:mod2} that accounts for the present day dark energy density. The excluded regions correspond to those described in the text around Figure~\ref{F:cons1}.\label{F:cons2}}
\end{figure}

The other constraints on theories with the zero temperature potential Eq.~\eqref{eq:mod2} are similar to the previous case, and the allowed parameter space is plotted in Figure~\ref{F:cons2}. As in the previous model, the zero temperature potential satisfies the dS swampland conjecture over all of the parameter space, and $\phi_1 \ll M_{\rm Pl}$.

\section{Observational signals and model building}\label{sec:signals}

The models that we consider can naturally lead to potentially observable signals, although these are not necessarily present. In this section we also address some model building issues, and discuss effects that depend on possible UV completions.

\subsubsection{Portal couplings and fifth force searches}\label{sec:portal}

Although the hidden sector that gives rise to Thermal Dark Energy does not have to couple directly to the visible sector, string theory UV completions suggest that portal couplings should exist at some level.  We will discuss the issue of sequestering between visible and hidden sectors in string models in Section~\ref{sec:UV}. In this section we focus on possible experimental signatures arising from portal interactions as well as their implications for fine-tuning. Such interactions are also constrained by their impact on the cosmological evolution of the hidden and visible sectors.

Limiting ourselves to hidden sectors without a U(1) gauge factor, the only renormalisible portal operators between $\phi$ and the visible sector involve the SM Higgs:\footnote{If there is a hidden sector U(1), kinetic mixing of this with the SM U(1) is also renormalisible.}
\be \label{eq:hp}
\mathcal{L} \supset - \left( A \phi + g \phi^2 \right) \left|H\right|^2 ~.
\ee
The experimental constraints on the linear coupling $A$ are far stronger than on the quadratic interaction $g$. It is most plausible that a non-zero value of $A$ will be present for the model of Section~\ref{sec:mod2}, since in this case any quantum numbers carried by $\phi$ are already broken in its own potential and cannot protect against a linear coupling to $\left|H\right|^2$. In contrast, in the model of Section~\ref{sec:mod} $\phi$ could carry a conserved quantum number, e.g. this could be a $\mathrm{Z}_2$ symmetry or in a straightforward extension $\phi$ could be a complex scalar and carry a gauge charge, which would forbid a linear interaction. On the other hand, $A$ is not necessarily zero even in this model, for example $\phi$ might be charged under a global symmetry, which could be violated by Planck scale effects \cite{Kamionkowski:1992mf}.

For the hidden sector scalar masses of interest, the strongest limits on the coupling $A$  are from experimental searches for fifth forces. The interaction Eq.~\eqref{eq:hp} leads to a Yukawa interaction between SM states, which modifies the potential between two objects separated by a distance $r$ to
\be \label{eq:yuk}
V_{ij} = -\frac{G m_i m_j}{r} \left(1+ \alpha_{ij} e^{-r/l} \right)  ~,
\ee
where $l = 1/m_{\rm phys}$ is set by the mass of the new scalar, and the coupling constants $\alpha_{ij}$ depend on the interactions of the new particle and the compositions of the bodies $i$ and $j$. 

\begin{figure}[t!]
\begin{center}
\includegraphics[scale=0.405]{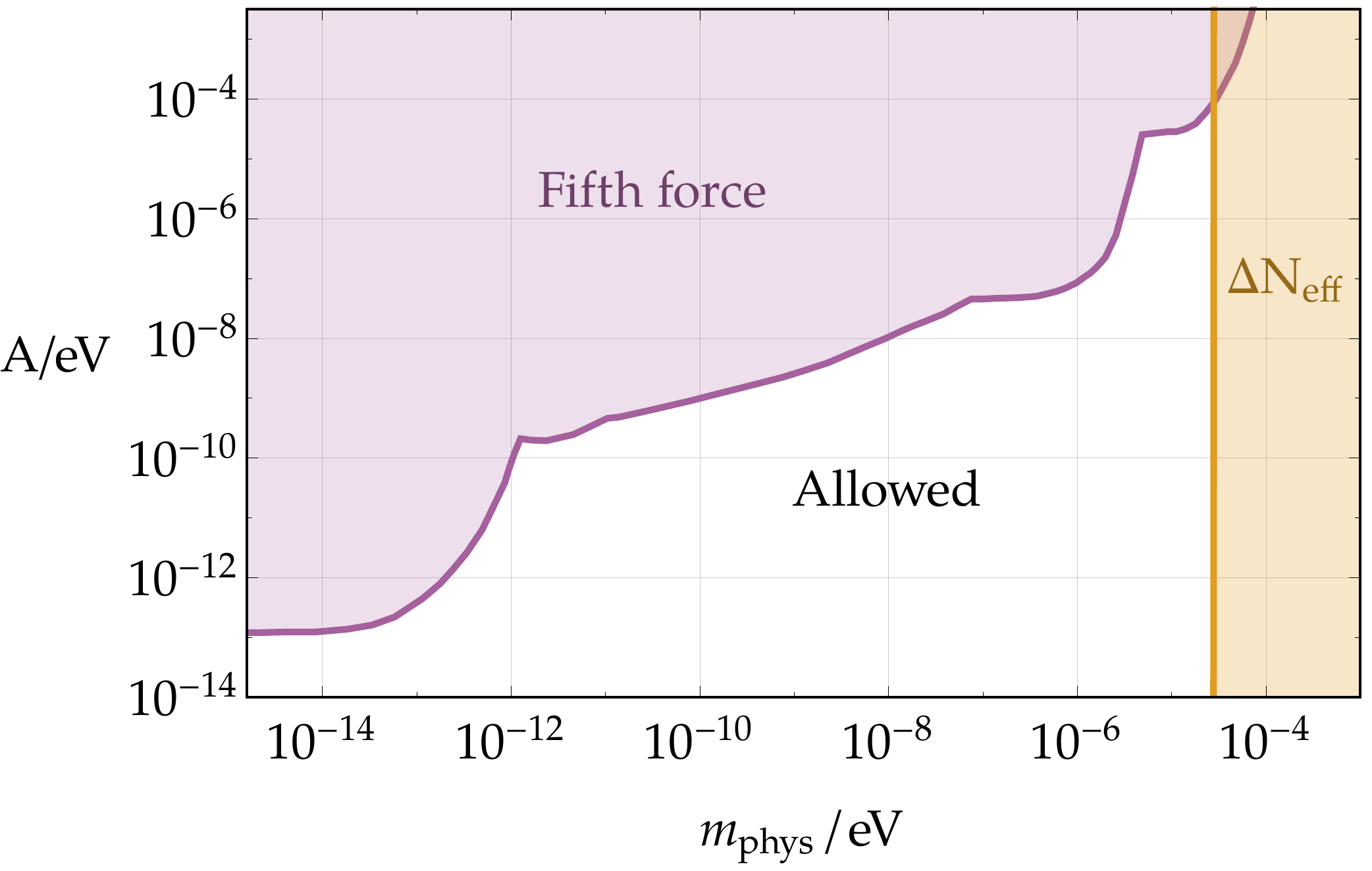}
\end{center}
\caption{The experimental bounds from fifth force searches on the interaction, with coupling constant $A$  defined by Eq.~\eqref{eq:hp}, of a new light scalar with the Standard Model Higgs as a function of the scalar's mass. The constraint remains constant at smaller values of $m_{\phi}$ than are plotted.  We also indicate the region of scalar masses that is excluded by constraints on the hidden sector temperature from $\Delta N_{\rm eff}$ measurements in our model, for the case that the scalar interacts with a single hidden sector Dirac fermion via a Yukawa coupling of $1$.  \label{F:ff1}}
\end{figure}

Spectacularly precise experiments have searched for the effects of a new long range Yukawa interaction. Following \cite{Piazza:2010ye}, null results from the searches \cite{Spero:1980zz,Hoskins:1985tn,Wagner:2012ui,Berge:2017ovy} can be interpreted as limits on $A$, and we plot the constraints obtained  in Figure~\ref{F:ff1}. In this plot $m_{\rm phys}$ is the physical mass of $\phi$ excitations around the metastable minimum defined in Eq.~\eqref{eq:mphys}, which does not coincide with the Lagrangian parameter $m_{\phi}$. For a model with $\phi$ coupled to a single Dirac fermion the bounds on $\Delta N_{\rm eff}$ mean that 
\be
m_{\rm phys} < 2.8 \times 10^{-5}~\eV~.
\ee

The coupling $A$ leads to logarithmically divergent contributions to the mass of $\phi$ from loop diagrams involving the SM Higgs. Provided $m_{\phi} \gtrsim A$, as is the case over large parts of the non-excluded parameter space in our model, these do not necessitate fine tuning the mass of $\phi$. However, $A$ also induces a quadratically divergent linear field shift of $\phi$ of the form $\mathcal{L} \sim A \phi \Lambda_{\rm UV}^2$, where $\Lambda_{\rm UV}$ is a UV scale that cuts off the visible sector loops \cite{Piazza:2010ye}. For this not to affect the dynamics of our model we need $A \Lambda_{\rm UV}^2 \ll {T_{\rm h}^0}^3$. This is a strong constraint, and values of $A$ that could be experimentally observed require that this contribution to $\phi$'s potential is fine tuned. However, given the uncertainty surrounding the solution of the SM electroweak hierarchy problem we remain agnostic on whether this issue should be interpreted as forbidding a significant value of $A$.

In contrast to $A$, fifth force experiments only weakly constrain the interaction that is quadratic in $\phi$ in Eq.~\eqref{eq:hp}, and comparable limits arise from the effects of the small shifts in the SM parameters induced by a $\left<\phi^2 \right>$ background in the early Universe \cite{Stadnik:2014tta,Stadnik:2015kia}. However, bounds from both of these sources turn out to be negligible in the models that we consider. Instead, since energy is transfered from the visible to hidden sector via this interaction, a stronger limit on the coupling $g$ comes from requiring that the hidden sector remains sufficiently cold that constraints on $\Delta N_{\rm eff}$ are evaded. The maximum relative energy transfer happens shortly after visible sector electroweak symmetry breaking when the SM Higgs gets a VEV and can decay directly to $\phi$, but before the visible sector temperature is small enough that the Higgs abundance is Boltzmann suppressed. For the hidden sector temperature not to change by more than a factor of $\mathcal{O}\left(1\right)$ we need $g < 10^{-10} \xi_{\rm h}$ \cite{Fairbairn:2019xog}. Additionally, loop diagrams involving $g$ lead to radiative corrections to $m_{\rm \phi}^2 \sim g \Lambda_{\rm UV}^2$, where $\Lambda_{\rm UV}$ is a UV cutoff in the visible sector. Consequently, unless it is tiny, a non-zero coupling $g$ may necessitate that $\phi$'s potential is fine tuned.

As well as the renormalisible portal interactions in Eq.~\eqref{eq:hp}, $\phi$ can interact with the visible sector through non-renormalisable interactions of the form
\be \label{eq:nonrn}
\mathcal{L} \supset \kappa_i \phi~ \mathcal{O}_{{\rm SM}i}  ~,
\ee
where $\kappa$ has mass dimension $-1$ and $\mathcal{O}_{{\rm SM}i}$ is an operator in the SM Lagrangian. For new scalars with a mass $\lesssim $meV fifth force limits require that $\kappa \lesssim M_{\rm Pl}^{-1}$ , i.e. the interactions must be weaker than gravitational strength.

The theoretical predictions for the patterns of $\{ \kappa_i \}$ in particular high energy theories and the phenomenology of the induced interactions have been studied extensively (see e.g. \cite{Damour:1994zq,Adelberger:2003zx,Damour:2010rp}). The values of the couplings $\kappa_i$ are extremely dependent on a model's UV completion, and they also evolve during renormalisation group (RG) flow between the UV scale and the low scale relevant to experimental searches. Similarly to the Higgs portal coupling that is linear in $\phi$, such interactions are most plausible in the model of Section~\ref{sec:mod2}, although they could also be present in the model of Section~\ref{sec:mod}.

Due to their UV dependence, we do not commit to a particular patten of $\{\kappa_i \}$. Instead we focus of the coupling to gluons, which is usually parameterised as
\be \label{eq:defdg}
\mathcal{L} \supset d_g \frac{\beta_3}{\sqrt{2} g_3 M_{\rm Pl}} \phi~ G_{\mu \nu} G^{\mu \nu} ~,
\ee
where $M_{\rm Pl}$ is the reduced Planck mass; and  $\beta_3$ is the beta function, $g_3$ is the gauge coupling, and $G_{\mu \nu}$ is the field strength of QCD. The coupling of $\phi$ with gluons is one of the interactions in Eq.~\eqref{eq:nonrn} that is most strongly experimentally constrained, and it is also often enhanced by a factor of $\sim 40$ relative to the other couplings by the RG flow from the string scale \cite{Damour:2010rp}. 

\begin{figure}[t!]
\begin{center}
\includegraphics[scale=0.39]{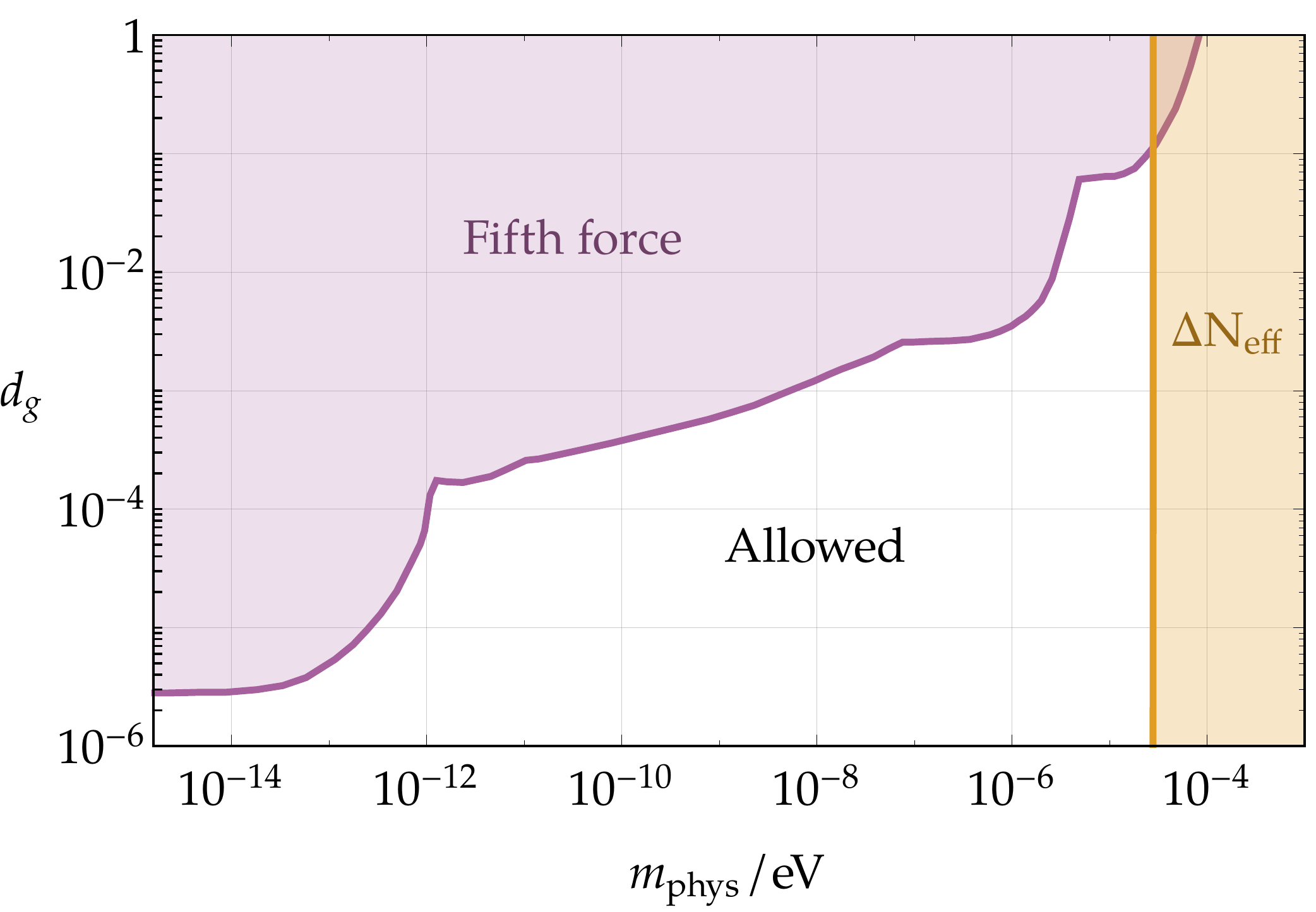}
\end{center}
\caption{The experimental bounds from fifth force searches on the coupling of a new light scalar to gluons via the interaction Eq.~\eqref{eq:defdg} with strength $d_g$. The constraint on the scalar mass from $\Delta N_{\rm eff}$ in our model is plotted similarly to as in Figure~\ref{F:ff1}. \label{F:ff2}}
\end{figure}

In Figure~\ref{F:ff2} we recast current fifth force constraints as bounds on $d_g$. Again these limits are on the physical mass of $\phi$ around the metastable minimum. Over a large part of the viable parameter space the bounds require interaction strengths that are not too much smaller than $M_{\rm Pl}^{-1}$.  Given the difficulty in finding string models in which the interactions in Eq.~\eqref{eq:nonrn} are significantly weaker than $M_{\rm Pl}^{-1}$ (\cite{Anisimov:2002az, Kachru:2006em, Berg:2010ha, Doran:2002bc,Acharya:2018deu,Hertzberg:2018suv, Heckman:2019bzm}), it is encouraging that such couplings are both not excluded and are also sufficiently close to current limits that they could plausibly be detected in upcoming experiments. 

On the other hand, such portal couplings have the downside of inducing loop diagrams that produce a $\phi$ tadpole, which is parametrically $\mathcal{L} \supset \kappa_i \Lambda_{\rm UV}^4 \phi$. This necessitates fine tuning in the hidden sector since $\kappa_i \Lambda_{\rm UV}^4 \gg {T_{\rm h}^0}^3$ for $\Lambda_{\rm UV} \gtrsim \TeV$. Despite this, as before, given the uncertain nature and solution of the SM hierarchy problem we do not regard fine tuning as a definitive problem, and fifth force experiments remain an interesting route to detecting $\phi$.

\subsubsection{Cosmological observables}

Theories of Thermal Dark Energy could also leave a detectable imprint on cosmological observables. One potential signal is a dark energy density that changes with time. However, for the simple potentials that we have studied this effect is actually negligible for a hidden sector that explains the present day energy density: In the model with a quartic interaction the metastable minimum remains fixed at $\phi=0$, and the induced dark energy density is not time dependent, until the temperature drops sufficiently low such that vacuum decay occurs. In the model of Section~\ref{sec:mod2} the location of the metastable minimum is time dependent via Eq.~\eqref{eq:min2}. Despite this, the change in the induced vacuum energy as the hidden sector temperature varies,
\be
\rho_{\rm DE} = \frac{m_{\phi}^2 \phi_1^2}{2} \left(1 - \frac{m_{\phi}^2}{2bT_{\rm h}^2}  \right) ~,
\ee
is negligible for models that account for the observed dark energy density, since the hidden sector temperature must satisfy Eq.~\eqref{eq:mod2cond}. An observationally significant time dependence might be possible in theories with more complex potentials, although we do not investigate this possibility further here.

Searches for beyond the SM contributions to $\Delta N_{\rm eff}$ are a more promising route to an detectable signal. In Figures~\ref{F:cons1} and \ref{F:cons2} a substantial proportion of the viable model space has a hidden sector temperature corresponding to a contribution to $\Delta N_{\rm eff}$ that is not far below the present limits, and which is within the reach of future measurements with improved sensitivities. Fairly high hidden sector temperatures are also beneficial from a UV perspective because they lead to relatively large physical $\phi$ masses, which relax fifth force constraints on portal couplings. Moreover, models with reasonably large values of $m_{\phi}$ have the appealing feature of not needing extremely small quartic couplings $\lambda$ (see Eq.~\eqref{eq:rhode}), and the hidden sector temperature must be relatively high for a metastable minimum to exist in such theories.  The increased number of relativistic degrees of freedom, parameterised by $\Delta N_{\rm eff}$ may also help alleviate the $H_0$ tension \cite{Wyman:2013lza}.

\subsubsection{Model building}

The zero temperature potentials that we have considered so far (Eqs.~\eqref{eq:vphi0} and ~\eqref{eq:mod2}) are extremely fine tuned. For it to source a significant amount of Thermal Dark Energy, $\phi$ must be strongly coupled to some additional hidden sector states. Then divergent loops of these will tend to generate a $\phi^4$ quartic coupling that is of $\mathcal{O}\left(1\right)$, which would destroy the required phenomenology in both of the potentials that we have considered. Further, the mass of $\phi$ is expected to be quadratically sensitive to a UV scale, analogously to the SM Higgs.\footnote{If $\phi$'s mass is sensitive to the Planck mass the required tuning is 1 part in $\sim 10^{66}$, and this is correspondingly reduced if the UV scale is lower.}

However, there are models of Thermal Dark Energy with all of the required phenomenological properties that need no fine tuning at all.\footnote{We do not speculate on the fine tuning required for the magnitude of the vacuum energy of the zero temperature potential to be sufficiently small for a viable theory.} As an example we show how this can be achieved using supersymmetry. Since it is necessarily weakly coupled to the visible sector, it is plausible that the hidden sector could have a supersymmetry breaking scale that is far smaller than the visible sector, and this is also reasonable from a string theory perspective as we discuss in Section~\ref{sec:UV}. 

A slight variant of the model in Section~\ref{sec:mod2} can easily be made supersymmetric. We suppose that the hidden sector contains two chiral superfields: $\Phi$, which has scalar component $\phi$ and fermion component $\xi$, and $\Psi$, which has scalar component $\chi$ and fermion component $\psi$. The theory's superpotential is fixed to 
\be \label{eq:W}
W =  \left( m_{\psi} -\Phi \right)  \Psi^2 ~,
\ee
and the \kahler potential is taken to be canonical. We assume that there are soft supersymmetry breaking scalar mass terms 
\be \label{eq:soft}
V_{\rm soft} = m_{\phi}^2 \left|\phi\right|^2 + m_{\chi}^2 \left|\chi\right|^2 ~,
\ee
which satisfy $m_{\psi} \gg m_{\phi}, m_{\chi}$. Supersymmetry protects against radiative corrections so this mass hierarchy is stable, as well as preventing any additional terms being generated in the superpotential.

The scalar potential corresponding to Eqs.~\eqref{eq:W} and \eqref{eq:soft} is
\be
V = \left| \chi \right|^4 + \left| 2 m_{\psi} \chi - 2 \phi \chi  \right|^2 + m_{\phi}^2 \left|\phi\right|^2 + m_{\chi}^2 \left|\chi\right|^2 ~,
\ee
and the Lagrangian contains fermion mass terms 
\be
\mathcal{L} \supset -\left( m_{\psi} - \left<\phi\right> \right) \psi^2 + 2 \left<\chi\right> \xi \psi ~.
\ee
At zero temperature the scalar potential is minimised at $\phi = \chi =0$.

Finite temperature effects break supersymmetry \cite{Girardello:1980vv}, as can be seen from the different ways that fermions and scalars contribute to the thermal potential in Eq.~\eqref{eq:thermpot}. As usual, the thermal potential favours parts of field space that minimise the masses induced by scalar VEVs. At large temperatures $T_{\rm h} \gg m_{\psi}$ the thermal potential has a global minimum at $\phi = m_{\psi}$ (with $\chi = 0$). For $m_{\phi}<T_{\rm h}<m_{\psi}$ this is no longer the global minimum, which is instead close to $\phi=0$, however it remains as a local minimum. Similarly to in the previous models the metastable minimum at $\phi = m_{\psi}$ leads to a contribution to the dark energy of 
\be
\rho_{\rm DE} = \frac{1}{2} m_{\psi}^2 m_{\phi}^2 ~.
\ee
This is analogous to the model in Section~\ref{sec:mod2} with $m_{\psi}$ replacing $\phi_1$. As before there is significant region of viable parameter space that can explain the present day dark energy density.

Finally, we address another model building issue. In the theory of Section~\ref{sec:mod} it is reasonable that there are additional fermions that are massless in the absence of a $\phi$ VEV, since $\phi$ can carry a conserved charge. In the model in Section~\ref{sec:mod2} a massless fermion is more unusual, since the potential includes terms $V \supset m_{\phi}^2 \phi_0 \phi$ and $ V \supset \phi \bar{\psi} \psi$ so there is no symmetry reason to forbid a fermion mass term $m_{f} \bar{\psi} \psi$. However, the addition of an explicit fermion mass does not destroy the dynamics provided $m_{\psi} \lesssim \phi_1$ (and viable models are even possible if $m_f \gg \phi_1$). In this case the thermal potential has a metastable minimum at $\phi = -m_f$ for sufficiently large hidden sector temperatures, and the dynamics remain the same as before with only minor modifications to the formula. In such theories a small quartic coupling $\lambda \phi^4$ also does not have a significant effect provided that $m_{\phi}/\sqrt{\lambda} \ll \phi_1 $.

\subsubsection{Sequestering and UV completions} \label{sec:UV}

If fine tuning is to be avoided, $\phi$'s potential must also be stable against corrections from loops of states at the scale of the theory's UV completion, which for us is typically the string scale, and loops of visible sector states interacting with the hidden sector gravitationally. Although it is not the only way that $\phi$'s potential can be protected, we will continue to use the supersymmetric hidden sector described previously as an example in which such effects can be discussed.

There is an unavoidable interaction between the hidden sector states and visible sector matter, or any other states in the theory, from exchange of gravitons. This leads to a contribution to $\phi$'s mass that is parametrically
\be
\Delta m_{\phi}^2 \sim \frac{1}{(4\pi)^6} \frac{M^6}{M_{\rm Pl}^4} ~,
\ee
where $M$ is the mass of the e.g. visible sector states (assuming an $\mathcal{O}\left(1\right)$ coupling between $\phi$ and the hidden sector matter fields) \cite{Burgess:2004ib}. For $m_{\rm SM} \lesssim 10^{5}~{\rm TeV}$ these contributions do not disrupt the hidden sector dynamics in the parameter space that we consider.

If the hidden sector hierarchies are protected by supersymmetry, the condition that loops of states at the string scale do not necessitate fine tuning is equivalent to demanding that the scale of hidden sector supersymmetry breaking remains sufficiently low. Given that the scale of the visible sector soft terms is $\gtrsim \TeV$ the source of supersymmetry breaking, parameterised by an F-term $F_{\rm b}$, must couple more strongly to the visible sector than the hidden sector. 

From a supergravity perspective the hidden sector soft terms are generically expected to be
\be \label{eq:sugrasoft}
m_{\rm soft} \gtrsim \alpha_{\rm loop} m_{3/2} ~,
\ee
where $m_{3/2} \sim F_{\rm b}/M_{\rm Pl}$ is the gravitino mass, and $\alpha_{\rm loop}$ is a loop factor set by the strongest interactions of a hidden sector state (for example this is the magnitude of the soft terms induced by anomaly mediation \cite{Randall:1998uk,Giudice:1998xp,Bagger:1999rd}). Such soft masses would be too large to avoid fine tuning in the hidden sector.

However, if the hidden sector is sequestered from the source of supersymmetry breaking, e.g. via geometric separation in the extra dimensions,  its soft terms could be parameterically suppressed from the expectation Eq.~\eqref{eq:sugrasoft} \cite{Burgess:2010sy}. This may occur if the volume of the internal compact dimensions is large, so the string scale is significantly lower than the Planck mass. For example, considering a particular proposed form of moduli stabilisation, \cite{Aparicio:2014wxa} found that the scale of the soft terms can be  
\be \label{eq:seqsoft}
m_{\rm soft} = \epsilon m_{3/2} ~,
\ee
where $\epsilon \sim m_{3/2}/M_{\rm Pl}$ is a small parameter.

In typical string models the magnitude of the hidden sector soft terms is closely linked to the mechanism by which the moduli are stabilised, and is model dependent (see e.g. \cite{Blumenhagen:2009gk}). 
As we do not commit ourselves to a particular string UV completion, we simply note that if the visible sector soft terms are generated by low scale gauge mediation  $F_{\rm b} \gtrsim \left( 10^5~\GeV \right)^2$, so the gravitino mass is $m_{3/2} \gtrsim 10^{-8}~\GeV$, a suppression of the form of Eq.~\eqref{eq:seqsoft} with $\epsilon \lesssim 10^{-7}$ would be sufficient to protect the hidden sector hierarchies. 

Sequestering the hidden sector from the visible sector can also prevent string scale physics from generating portal couplings between the two that are excluded by fifth force searches. The experimental constraints on the portal interactions Eq.~\eqref{eq:nonrn} require $\kappa_i \lesssim M_{\rm Pl}^{-1}$, which is in tension with the naive expectation that UV modes will produce $\kappa_i \sim M_{\rm Pl}^{-1}$. However, in sequestered models the interactions of $\phi$ with the visible sector may be suppressed from Planck scale by some (model-dependent) power of the volume modulus \cite{Randall:1998uk, Kachru:2007xp, Burgess:2010sy, Cicoli:2012tz}. Given the range of physical $\phi$ masses that we are interested in, only a fairly mild reduction in the $\kappa_i$ from $M_{\rm Pl}^{-1}$ is necessary, which appears achievable in explicit models.

\subsubsection{Thermal dark energy during other eras}

Although we have focused on models that account for the present day dark energy, there could also be hidden sectors that give a Thermal Dark Energy contribution at early times and have subsequently decayed to the global minimum of their potential. Indeed in a string theory context it is plausible that there could be multiple hidden sectors at differing mass scales and temperatures, each sourcing a component of dark energy that disappears at a different time.\footnote{Previously considered models of Thermal Inflation \cite{Lyth:1995hj,Lyth:1995ka} correspond to the visible sector itself sourcing a component of Thermal Dark Energy at scales $\sim \TeV$ that dominates the energy density of the Universe for a significant number of e-folds.} Depending on the scales and interactions of a hidden sector, the resulting dynamics can lead to unusual phenomenology and signals, and could also dramatically modify the relic abundances of visible and hidden sector relics. We leave the interesting task of performing a detailed analysis of this scenario to future work, and instead here we simply point out a few preliminary considerations and possibilities.

The cosmological impact of such a hidden sector depends on the scale at which it decays to the global minimum of its potential, its temperature hierarchy relative to the visible sector, whether it has any interactions with the visible sector, and the maximum ratio between the density of the dark energy sourced and the radiation energy density that is reached before the metastable vacuum decays. For example, for the quartic potential of Section~\ref{sec:mod} the maximum ratio, reached immediately prior to decay to the global minimum, is approximately
\be
\frac{\rho_{\rm DE}}{\rho_{\rm rad}} \sim \frac{\xi_{\rm h}^4}{\lambda} ~.
\ee
The corresponding number of e-folds of accelerated expansion is given, analogously to Eq.~\eqref{eq:nefold}, by
\be
N_{\rm DE} \sim \log\left(\frac{\xi_{\rm h}}{\lambda^{1/4}}\right) \nonumber ~.
\ee
If an era of dark energy domination happens at sufficiently early times, and the hidden states are sufficiently heavy, it is possible that the hidden sector energy density could be transfered to the visible sector after the metastable minimum decays. Otherwise limits on the hidden sector energy density from $\Delta N_{\rm eff}$, and the abundance of stable hidden sector relics, will be relevant.  It is also possible that the early dark energy does not dominate the total energy density of the Universe.

There is a long standing discrepancy between cosmological and astrophysical measurements of the Hubble parameter (see e.g. \cite{Feeney:2017sgx,Freedman:2017yms,DiValentino:2017zyq} and references therein). It has been shown that this could be explained by an additional contribution to the dark energy that is present at early times and after $z \sim 3000$ redshifts as radiation (or faster) \cite{Karwal:2016vyq,Poulin:2018cxd}. Such dynamics could naturally arise from a period of early Thermal Dark Energy, if a hidden sector transitions to the minimum of its zero temperature potential at this time.

As discussed in Section~\ref{sec:decays}, a hidden sector that sourced an early Thermal Dark Energy component will transition to its zero temperature minimum   
through the nucleation of critical bubbles, which subsequently expand and collide. This produces a background of stochastic gravitational waves, which could be detected in current and proposed experiments (for a recent review see e.g. \cite{Caprini:2015zlo}). The frequency of the gravitational wave signal is parametrically fixed by the value of the Hubble parameter at the time of the transition, appropriately redshifted, and the amplitude of the signal is determined by the energy released and the properties of the transition, such as the time it takes to complete. Gravitational wave signals from models of Thermal Inflation at a temperature around the TeV scale have been considered in \cite{Easther:2008sx}. However, given the possibility of Thermal Dark Energy in hidden sectors at a wide range of mass scales, signals at many different frequencies are possible.

\section{Discussion} \label{sec:discuss}

We have shown that Thermal Dark Energy is a viable route to an effective `cosmological constant' that can persist to late times. It is therefore a possible explanation for the present day accelerated expansion of the Universe. This is despite the theory's zero temperature potential having no de Sitter vacuum or slow roll directions.

We have considered two simple realisations of the scenario, one with a hidden Higgs-like quartic scalar potential, $V(\phi) = \lambda \phi^4 -\frac12 m_\phi^2 + C$, and the other with a string modulus like potential, $V(\phi)=\frac12 (\phi-\phi_1)^2$, both with a Yukawa-like coupling between $\phi$ and a hidden sector fermion.  Both theories are consistent with recent string swampland conjectures, assuming that the finite temperature potential need not satisfy the dS conjecture. This is reasonable, since the finite temperature potential of the SM Higgs already violates Eq.~\eqref{E:swamp} in the early Universe. Similarly to the model of Section~\ref{sec:mod}, prior to the EW phase transition finite temperature effects source a dS minimum around $H=0$.\footnote{For the SM Higgs radiation energy density dominates while it is in a thermally generated minimum, so it does not result in the expansion of the Universe accelerating.} 
 It would however be very interesting to study the status of the dS conjecture at finite temperature carefully, potentially applying the framework of \cite{Ooguri:2018wrx}.

While both a dS zero temperature vacuum and also viable theories of quintessence are difficult to construct within string theory, Thermal Dark Energy models that explain the present day dark energy density might arise naturally. The essential phenomenological ingredient of a low scale hidden sector, at a lower temperature than the visible sector CMB photons and still in internal thermal equilibrium today, seems reasonable given what is known about generic string compactifications. The condition that the hidden sector includes a scalar field with Higgs-like couplings to hidden matter also does not seem surprising. One unusual requirement is the need for a moderate hierarchy of scales, or a small dimensionless coupling, in the hidden sector. However, even this may be made technically natural. In particular, we have shown how Thermal Dark Energy can occur in a supersymmetric model for which the necessarily small scalar quartic self-coupling is protected against loops of hidden sector states  provided the hidden sector supersymmetry breaking soft terms are sufficiently small.

To avoid the need for fine tuning, $\phi$'s potential must also be robust against quantum corrections from loops of heavy states, for example associated to the string scale, which depends on the theory's UV completion. Taking the model with hidden sector supersymmetry as an example, there is hope that this can be addressed by sequestering the hidden sector from the source of supersymmetry breaking.\footnote{The issue of producing sufficiently small couplings, along with the need to protect $\phi$'s potential against radiative correction, would be resolved if $\phi$ were an axion field. However, from preliminary investigations it seems difficult to construct models in which finite temperature contributions to an axion's potential shift its VEV and generate a temperature dependent potential energy.} As discussed in Section~\ref{sec:UV}, the hidden sector soft terms must be $\lesssim 10^{-6}~\eV$, which is smaller than their naive supergravity estimated value $\sim m_{3/2}$. However, the required suppression seems reasonable given what is currently known about string compactifications, and certainly appears easier to achieve than the scales necessary for quintessence. Moreover, as Thermal Dark Energy decouples dark energy from the moduli stabilisation problem in string theory -- though all moduli must still be stabilised -- new model building avenues should be possible. Sequestering also helps to keep the hidden sector in thermal isolation from the visible sector, and it might explain why the hidden sector was reheated to a lower temperature than the visible sector.

Having studied the observational and model building constraints in Section~\ref{sec:param}, we have seen that both of the zero temperature scalar potentials that we considered have large regions of viable parameter space in which they can explain the present day dark energy. 

Over a significant part of the allowed parameter space, signals are possible in future experimental searches. For example, UV modes often generate non-renormalisible portal couplings between the hidden and visible sectors with interaction strengths that are of order $M_{\rm Pl}^{-1}$, although the couplings could be somewhat suppressed relative to this in sequestered string models. Meanwhile experimental searches for fifth forces constrain linear interactions of $\phi$ with the visible sector to be slightly weaker than Planckian in the mass range of interest. So, with an improvement in sensitivities, a future detection is plausible. We do however note that couplings that are large enough to be observed are likely to introduce quantum corrections to $\phi$'s potential that necessitate fine tuning. From a UV perspective, the relatively large physical masses of $\phi$ in the models that we have considered, $m_{\rm phys} \lesssim 10^{-5}~\eV$, are beneficial. As seen in Figure~\ref{F:ff2}, the fifth force constraints at these masses are far weaker than those on scalars with masses $\lesssim 10^{-12}~\eV$. Consequently, whereas in string theory it has proven difficult to obtain sufficient sequestering between quintessence and the visible sector for models to not already be excluded \cite{Anisimov:2002az, Kachru:2006em, Berg:2010ha, Doran:2002bc,Acharya:2018deu,Hertzberg:2018suv, Heckman:2019bzm}, the required suppression of couplings seems within reach in our models.

The hidden sector's contribution to $\Delta N_{\rm eff}$ is also often close to present limits and within reach of future observations.  Moreover, if there are several distinct Thermal Dark Energy sectors this can lead to additional cosmological effects including the $H_0$ tension; and sectors that have decayed to the zero temperature minimum of their potential naturally emit potentially observable gravitational wave signals since this involves a first order phase transition.

Finally, we have assumed that some unknown mechanism exactly cancels the zero temperature contributions to the vacuum energy from the SM, hidden sectors and any other new physics, with the view that is easier to explain zero than $10^{-120}M_{\rm Pl}^4$.

To conclude, Thermal Dark Energy appears to be a promising candidate to explain the current accelerated expansion of our Universe.  It will be important to study its potential signatures in more detail, and to try to embed the scenario into a UV complete theory of quantum gravity, such as string theory.

\section*{Acknowledgments}
It is our pleasure to thank Tommi Markkanen, Jamie Rogers, Gianmassimo Tasinato, Radu Tatar and Ivonne Zavala for helpful discussions.

\bibliography{refs}
\bibliographystyle{utphys}

\end{document}